\documentclass[aps,pra,twocolumn,groupedaddress,amsmath,amssymb,10pt]{revtex4-1}
\usepackage{hyperref}
\usepackage{graphicx,dsfont,amsthm}

\newcommand{\bra}[1]{\left\langle{#1}\right\vert}
\newcommand{\ket}[1]{\left\vert{#1}\right\rangle}

\newcommand{\norm}[1]{\bigl\|{#1}\bigr\|_1}
\newcommand{\epsbar}{\bar{\varepsilon}}

\newcommand{\overlap}[2]{\left<{#1}|{#2}\right>}
\newcommand{\proj}[1]{\left\vert{#1}\right\rangle \left\langle{#1}\right\vert}
\newcommand{\p}[1]{P_{\left\vert{#1}\right\rangle}}

\DeclareMathOperator{\tr}{tr}

\newcommand{\Prob}[1]{\mathrm{Prob}\left[#1\right]}
\newcommand{\dist}[1]{\frac{1}{2}\left|\left|#1\right|\right|_1}

\newcommand{\ketbra}[2]{{|#1\rangle\!\langle#2|}}

\newcommand{\eps}[1]{\varepsilon_{\mathrm{#1}}}

\begin{document}

\newtheorem{thm}{Theorem}
\newtheorem{lem}{Lemma}
\newtheorem{defi}{Definition}

\title{Min-entropy and quantum key distribution: non-zero key rates for ``small'' numbers of signals}
\author{Sylvia Bratzik}
\email{bratzik@thphy.uni-duesseldorf.de}
\author{Markus Mertz}
\author{Hermann Kampermann}
\author{Dagmar Bru{\ss}}
\affiliation{Institute for Theoretical Physics III, Heinrich-Heine-Universit\"at D\"usseldorf, 40225 D\"usseldorf, Germany.}

\date{\today}

\begin{abstract}
We calculate an achievable secret key rate for quantum key distribution with a finite number of signals, by evaluating the min-entropy explicitly. The min-entropy can be expressed in terms of the guessing probability, which we calculate for $d$-dimensional systems. We compare these key rates to previous approaches using the von Neumann entropy and find non-zero key rates for a smaller number of signals. Furthermore, we improve the secret key rates by modifying the parameter estimation step. Both improvements taken together lead to non-zero key rates for only $10^4-10^5$ signals. An interesting conclusion can also be drawn from the additivity of the min-entropy and its relation to the guessing probability: for a set of symmetric tensor product states the optimal minimum-error discrimination (MED) measurement is the optimal MED measurement on each subsystem. 
\end{abstract}


\maketitle

\section{Introduction}

Quantum key distribution (QKD) is the establishment of a random secure key between two authorized parties, Alice and Bob, which are connected with each other via a quantum and a classical channel \cite{BB84}. Qubits (e.g.\ photons) are distributed over this quantum channel, and in practical implementations the number of these particles is finite. Dealing with these finite resources, a new branch in quantum key distribution (QKD) emerged, the finite-key analysis. It investigates secure key rates, i.e.\ the ratio of a secure key length to the number of signals sent through the channel, in the non-asymptotic situation. The security of a finite key for a composable security definition \cite{Ben04,Unr04,Mue09} was proven in \cite{RenKoe,Ren05a,Kra05,Hay07,Sca08a,RenPhD}. It is important to notice that composability means that the key established by QKD can be used safely in any application such as one-time-pad encryption. For a review on practical QKD and its security, see e.g.\ Refs.\ \cite{Gisin2002,Sca09}.
 Calculations of finite key rates were done in \cite{Mey06,Sca08a,Cai09} and in \cite{She10} for $d$ dimensions. The relevance of finite QKD was shown in \cite{Has07}: practical implementations of QKD lead to a dramatically lower secure key rate in comparison to asymptotic theoretical predictions. 

The paper is organized as follows: in Sec.~\ref{sec:tomographic} we describe a general QKD protocol, in Sec.~\ref{sec:parameter} we review a bound for the statistical error in parameter estimation and show that former results on the secret key rate \cite{Sca08a} can be improved by considering a POVM with two outcomes. In Sec.~\ref{sec:privacy} we concentrate on quantifying the secret key length after privacy amplification. It was found in \cite{RenPhD,Sca08a} that the \textit{conditional min-entropy} gives an achievable upper bound on the secret key length. The calculation of the conditional min-entropy involves an optimization over a set of quantum states. A lower bound on the min-entropy by using the conditional von Neumann entropy was established in \cite{RenPhD,Sca08a}. This bound holds under the assumption of collective attacks, i.e.\ the state shared between Alice and Bob after Eve's interaction has tensor product structure. 
In Sec.~\ref{sec:guessing} we calculate the min-entropy explicitly by applying recent results on its operational meaning \cite{Sch08}. For the qubit case, we evaluate the min-entropy for the BB84-protocol \cite{BB84} via minimum-error discrimination (MED). For $d$-dimensional quantum systems, we calculate it for the generalized six-state-protocol \cite{Bru98,Bec99} via the square-root measurement.  In Sec.~\ref{sec:comparison} we compare the key rates via calculation of the min-entropy to the bound with the von Neumann entropy. We show that our approach gives positive key rates for a smaller number of signals compared to the von Neumann approach. Furthermore we compare our results in the $d$-dimensional case to the recent results in \cite{She10} for the mentioned bound. We conclude in Sec.~\ref{sec:conclusion}.

\section{Quantum key distribution protocol}
\label{sec:tomographic}

We consider an entanglement-based QKD scheme. In the following a description of the protocol will be provided.

1.\ \textit{Distribution:} Alice prepares $N$ maximally entangled states in dimension $d \times d$, where $d$ is the dimension of the Hilbert space of a subsystem:
\begin{equation}
\ket{\Phi_{00}}:=\frac{1}{\sqrt{d}}\sum_{x=0}^{d-1}\ket{xx},
\end{equation}
and sends the second particle to Bob. In the case of qubits, i.e.\ $d = 2$, the state is one of the four Bell-states $\ket{\Phi^+}$. After the distribution, they share $N$ entangled pairs, which we will denote by the state $\tilde{\rho}_{A^NB^N}$. Under the assumption of collective attacks, the state $\tilde{\rho}_{A^NB^N}$ is a tensor product state, i.e.\ $\tilde{\rho}_{A^NB^N}=\left(\rho_{AB} \right)^{\otimes N}$ \cite{Sca08a}. Alice and Bob can symmetrize the state $\rho_{AB}$ by applying a depolarizing map, leading to a $d^2$-dimensional Bell-diagonal state \cite{Ren05a,Kra05,She10}:
\begin{equation}
\rho_{AB}=\sum_{j,k=0}^{d-1}\lambda_{jk}\proj{\Phi_{jk}},
\label{eqn:belld}
\end{equation} 
where $\ket{\Phi_{jk}}=\frac{1}{\sqrt{d}}\sum_{s=0}^{d-1}\left(e^{\frac{2\pi i}{d}}\right)^{sk}\ket{s}\ket{(s+j) \mod d}$ are the generalized Bell states \cite{Ben93}. 
For $d=2$, the state $\rho_{AB}$ has the following form:
\begin{equation}
\rho_{AB}=\lambda_{00} \p{\Phi^+}+ \lambda_{01} \p{\Phi^-}+ \lambda_{10} \p{\Psi^+}+\lambda_{11} \p{\Psi^-},
\label{eqn:bell2}
\end{equation}
where $\p{\psi}=\proj{\psi}$, $\sum_{i,j} \lambda_{ij} =1$ and $\{\ket{\Phi^+},\ket{\Phi^-},\ket{\Psi^+},\ket{\Psi^-}\}$ is the Bell basis. 

For a specific depolarizing map, one can parametrize the state $\rho_{AB}$ by one parameter $Q$, which in the two-dimensional case is the \textit{quantum bit error rate} (QBER). The relation between $Q$ and $\lambda_{jk}$ will be explained in Sec.~\ref{sec:guessing}.

2.\ \textit{Encoding and measurement:} Both parties agree on an encoding, i.e.\ each quantum state is associated with a symbol from an alphabet. They perform a projection measurement in certain bases.

After this step Alice and Bob will share $N$ correlated pairs of \textit{d}its ($d$-letter systems).

3.\ \textit{Sifting:} In this step both parties announce for each qudit pair the encoding they have chosen. Depending on the protocol either they discard the data, when they differ, or they use them for parameter estimation. The bit string after this process has length $N-n'$, when $n'$ bits were discarded. 

4.\ \textit{Parameter estimation} (PE): 
Parameter estimation serves for estimating the error in the quantum channel by using measurements, in general a positive operator valued measure (POVM). The considered state is parametrized by the quantum bit error rate (QBER) $Q$ for $d=2$. To measure the QBER a
chosen POVM is used. Due to the finite number of signals ($m$ randomly chosen signals are used) the QBER cannot be detected perfectly. Therefore a quantification of the statistical error is needed. 

After parameter estimation the number of signal states is $n=N-n'-m$.

5.\ \textit{Error correction} (EC): In this step Alice and Bob want to eliminate the error in their classical data, which might be there because of eavesdropping. In order to reconcile their data, they have to communicate publicly. In this paper, we will use known results \cite{Sca08a} to account for the effect of error correction on the key.

6.\ \textit{Privacy amplification} (PA): During the key generation, information about the key might have been revealed to the eavesdropper. To reduce this information, Alice and Bob apply a randomly chosen hash function from a family of hash functions to their identical keys.
\section{Improved parameter estimation}
\label{sec:parameter}
Parameter estimation plays an important role in finite QKD protocols. Since one has a finite number of measurement outcomes, one needs an appropriate estimate for each parameter. In this section we first remind the reader of a method for parameter estimation used in \cite{Sca08,Sca08a,Cai09}. There, the parameters were estimated by different two-dimensional POVMs for different bases. We will then show that we can reach a better approximation, if we consider one specific POVM for the estimation of all parameters. The following theorem quantifies the unavoidable statistical errors in the estimated parameters.

\begin{thm}
\label{thm:1}
\cite{Sca08,Sca08a,Cai09} Let $\left\lbrace B_i\right\rbrace_{i=1}^{|\chi|}$ be a $|\chi|$-dimensional POVM, $\vec{\lambda}_m=\left(\lambda_m(1),\lambda_m(2),..,\lambda_m(|\chi|)\right)$ and $\vec{\lambda}_\infty=\left(\lambda_\infty(1),\lambda_\infty(2),..,\lambda_\infty(|\chi|)\right)$
 the probability distributions, with $\lambda(i)$ being the probability of outcome $B_i$. Here, the index $m$ stands for the $m$-fold independent application of the POVM on identical states $\rho$. Let now $\lambda_m:=\lambda_m(k)$, $\lambda_\infty:=\lambda_\infty(k)$ denote any $k$-th parameter. 
Then except with probability $\eps{PE}$ 
 \begin{equation}\label{PE}
  \dist{\lambda_m-\lambda_\infty}\leq \xi(\eps{PE},|\chi|,m),
 \end{equation}
\begin{equation}
\xi(\eps{PE},|\chi|,m):=\sqrt{\frac{\ln{\left(\frac{1}{\eps{PE}}\right)}+|\chi|\ln{(m+1)}}{8m}},
\label{eq:zeta}
\end{equation}
where $\left|\left|A\right|\right|_1=\tr \sqrt{A^\dagger A}$ and $\ln$ denotes the natural logarithm \footnote{The formula in \cite{Sca08,Sca08a,Cai09} was corrected in an erratum \cite{Cai09err}. The formula in Eq.~\eqref{eq:zeta} can be obtained by multiplying the corresponding formula in \cite{Cai09err} by $\frac{1}{2}$.}.
\end{thm}
\emph{Proof:} See Appendix.

To clarify the influence of different choices of POVMs on secure key rates we consider a protocol, where Alice and Bob share a state, which can be parametrized by $n_{\mathrm{PE}}$ parameters. We choose the variables of the estimation of each parameter in a symmetric way. That means $\eps{PE_i}=\frac{\eps{PE}}{n_{\mathrm{PE}}}, |\chi|_i=|\chi|, m_i=\frac{m}{n_{\mathrm{PE}}}$ for all $i\in \{1,...,n_{\mathrm{PE}}\}$, such that the constraints $\sum_{i=1}^{n_{\mathrm{PE}}} \eps{PE_i}=\eps{PE}$ and $\sum_{i=1}^{n_{\mathrm{PE}}} m_i=m$ are fulfilled. 

In previous works \cite{Sca08a,Cai09} each parameter is estimated by an individual two-dimensional POVM (in the following we will use IPOVM as an abbreviation for this approach);
e.g.\ for the BB84 protocol, we have two parameters (error rates in two bases) to estimate. Then we need two POVMs, where each of them has two outcomes which correspond to ``Alice and Bob \textit{do have} the same measurement outcome" and  ``Alice and Bob \textit{do not have} the same measurement outcome" in their respective measurement basis. This leads to $\xi(\frac{\eps{PE}}{2},2,\frac{m}{2})$.
Generally for states determined by $n_{\mathrm{PE}}$  we get $\xi(\frac{\eps{PE}}{n_{\mathrm{PE}}},2,\frac{m}{n_{\mathrm{PE}}})$ for each parameter.

Concerning secure key rates we can improve this method by considering a common POVM with $n_{\mathrm{PE}}+1$ measurement outcomes (CPOVM approach). This means for example for the BB84 protocol that we use a POVM with $3$ outcomes, where two of them correspond to  ``Alice and Bob \textit{do not have} the same measurement outcome" in each of the two bases and one corresponds to the completeness of the POVM. Then, the estimation of each parameter will be represented by $\xi(\eps{PE},3,m)$. In general for $n_{\mathrm{PE}}+1$-dimensional systems the deviation from the perfect parameter (see Eq.~\eqref{PE}) is given by $\xi(\eps{PE},n_{\mathrm{PE}}+1,m)$. The improvement is due to the fact that in Eq.~\eqref{PE} the trace distance is only bounded by $\xi(\eps{PE},|\chi|,m)$ and that the parameters according to the CPOVM approach lead to a smaller bound than the IPOVM approach. The results of an explicit calculation of the key rates will be provided in the last section.

\section{Privacy amplification and the $\epsbar$-smooth min-entropy}
\label{sec:privacy}
In this section, we will present some results about the min-entropy. Starting from the connection of the min-entropy to the secure key length after the \textit{privacy amplification} step, we review the relation of the min-entropy to the guessing probability given in \cite{Sch08}. 

\subsection{The $\epsbar$-smooth min-entropy and the secure key length $\ell$}
The $\epsbar$-smooth conditional min-entropy provides an upper bound for the secure key length $\ell$ after the \textit{privacy amplification} step \cite{RenPhD}:
\begin{equation}
 \ell \lessapprox H_{\min}^{\epsbar}(\rho_{XE}^{\otimes n}|E^n),
\label{eq:keylength}
\end{equation}
where $\rho_{XE}=\sum_{x=0}^{d-1} p_x \proj{x}\otimes \rho_E^x$ is a \textit{classical-quantum}-state, which Alice and the eavesdropper Eve share after error correction. Here, $X$ is Alice's random variable with values $x\in\{0,...,d-1\}$, where $d$ is the dimension of the quantum system. The eavesdropper holds a quantum state $\rho_E^x$, which is correlated with the random variable $X$. The symbol $E^n$ denotes the eavesdropper's system. The parameter $n$ is the number of signals after sifting and parameter estimation, i.e.\ $n=N-n'-m$.

In the following we will denote the state $\rho_{XE}$ as a \textit{single-signal state}, i.e.\ following the description above, both parties share one single state ($n=1$). Otherwise, the state will be denoted by $\rho_{XE}^{\otimes n}$, if it has tensor product structure. We treat collective attacks, as the state shared between Alice and Eve has in this case tensor product structure. Collective attacks \cite{Bih97,*Biham2002} are those attacks where the eavesdropper is restricted to interact with each of the signals separately, i.e.\ by attaching an auxiliary system and performing unitary transformations. In \cite{Kra05,Ren05a} it was shown that it suffices to consider a convex combination of product states, when analyzing the full security of QKD protocols. However, it does not follow that we can consider w.l.o.g.\ a product state.

We recall the definition of the $\epsbar$-smooth min-entropy:
\begin{defi}[$\epsbar$-smooth min-entropy \cite{RenPhD}]
Let $\bar{\rho}_{XE} \in \mathcal B^{\epsbar/2}(\rho_{XE}):=\{\bar{\rho}_{XE} \geq 0:\norm{\bar{\rho}_{XE}-\rho_{XE}}\leq \epsbar \}$. The $\epsbar$-smooth min-entropy is defined as:
\begin{equation}
H_{\min}^{\epsbar}(\rho_{XE}|E):=\sup_{\bar{\rho}_{XE}}H_{\min}(\bar{\rho}_{XE}|E),
\label{eq:min}
\end{equation}
with
\begin{equation}
H_{\min}(\bar{\rho}_{XE}|E)=\sup_{\sigma_E}\left[-\log_2 \left( \min \lambda: \lambda \cdot \mathds{1}_X \otimes \sigma_E \geq \bar{\rho}_{XE}  \right)\right].
\label{eq:min2}
\end{equation}
\end{defi}

The optimization in Eq.~\eqref{eq:min} is done over the states $\bar{\rho}_{XE}$ in the $\epsbar$-environment of $\rho_{XE}$, whereas the optimization in Eq.~\eqref{eq:min2} is over all states $\sigma_E$ . 

\subsection{The min-entropy and the guessing probability}

The evaluation in Eq.~\eqref{eq:min} is a convex optimization problem. It was shown in \cite{Sch08}, that the min-entropy can be rewritten as the negative logarithm of the optimal guessing probability $\text{p}_\text{guess}$ :
\begin{equation}
H_{\min}(\rho_{XE}|E)=-\log_2 \text{p}_\text{guess},
\label{eq:min-guess}
\end{equation}
where 
\begin{equation}
\text{p}_\text{guess}\equiv\text{p}_\text{guess}(X|E):=  \max_{ \{E_E^{x} \}} \sum_{x=0}^{d-1}p_x \text{tr}(E_E^x \rho_E^x).
\label{guess}
\end{equation}
Here it was used that the initial state $\rho_{XE}$ is a classical-quantum state (see above), which is shared between Alice and Eve, the eavesdropper. The set $\{E_E^x \}$ denotes the POVM elements of Eve, which she uses in order to distinguish her nonorthogonal ancilla states $\rho_E^{x}$. If she could perfectly discriminate them, she would know the value of Alice's random variable $X$ and therefore the content of the secret key.

\section{Evaluation of the guessing probability}
\label{sec:guessing}
In this section we will present an explicit calculation of the guessing probability in Eq.~(\ref{guess}) for $d$-dimensional quantum systems for the generalized six-state-protocol via \textit{square-root measurement} (see e.g.\ \cite{Hol78,Hau94,Ban97,Eld01,Che00,Bar01}), and for qubit-systems ($d=2$) for the BB84-protocol via \textit{minimum-error discrimination} \cite{Hol73,Yuen75,Hel76,Ban97,Her04}. The problem of distinguishing two mixed quantum states with minimum error was solved by Helstr\o{}m \cite{Hel76}, but for more states it becomes more involved. For quantum states with a certain symmetry, optimal measurements were found (see e.g.\ \cite{Bar01}), whereas for arbitrary states only bounds exist \cite{Qiu10}. Finally, we draw a conclusion from the additivity of the min-entropy for tensor product states: for a set of symmetric tensor product states the optimal minimum-error discrimination (MED) measurement is the optimal MED measurement on the subsystems.


\subsection{Generalized six-state-protocol for $d$-dimensional quantum systems}

In this part we consider a $(d+1)$-bases protocol, which was introduced in \cite{Bechmann2000,Bru02,Cer02}. It is a generalization of the six-state protocol \cite{Bru98,Bec99}. We further assume a collective eavesdropping attack. Due to symmetrizations \cite{Ren05a} the eavesdropper is forced to introduce the same error in each measurement basis. This symmetrization leads to the following Bell-diagonal state shared between Alice and Bob (see Section~\ref{sec:tomographic}):
\begin{equation}
 \rho_{AB}=(\beta_0 -\beta_1)\ket{\Phi_{00}}\bra{\Phi_{00}}+\frac{\beta_1}{d}\mathds{1}_{d^2},
\label{eqn:bell}
\end{equation}
with $\beta_0+(d-1)\beta_1=1$, $0 \leq \beta_1 <\frac{1}{d}<\beta_0\leq 1$ and $\mathds{1}_{d^2}$ being the identity matrix of size $d^2$. Note that this form is equal to the one considered in \cite{Bru03,Lia03}.
The parameter $\beta_0$ can be seen as the probability that both get the same output, whereas $\beta_1$ denotes the  probability that they get a particular other one. The error rate $Q$ is given by $Q:=1-\beta_0=(d-1)\beta_1$; for $d=2$, $Q$ is the quantum bit error rate $\beta_1$. The state in Eq.~\eqref{eqn:bell} can be recovered from Eq.~\eqref{eqn:belld} by setting $\lambda_{00}=1-\frac{d+1}{d}(1-\beta_0)$ and all other $\lambda_{jk}=\frac{(1-\beta_0)}{d(d-1)}=\frac{\beta_1}{d}$.
 
We assume that Eve holds a purification $\ket{\psi_{ABE}}$. Eve's reduced state is \cite{Bru03}
\begin{equation}
\rho_E= \frac{1}{d}\left(\beta_0\sum_{x=0}^{d-1} \ket{E_{xx}}\bra{E_{xx}}+\beta_1 \sum_{x,y \atop y \neq x} \ket{E_{xy}}\bra{E_{xy}}  \right),
\label{eqn:rhoE}
\end{equation}
and we define the normalized states $\rho_E^x$ as:
\begin{equation}
\rho_E^x:=\beta_0\ket{E_{xx}}\bra{E_{xx}}+\beta_1 \sum_{y \neq x} \ket{E_{xy}}\bra{E_{xy}},
\label{eqn:rhoEx}
\end{equation}
such that Eve's state is given by  $\rho_E=\frac{1}{d}\sum_x \rho_E^x$. Eve's ancilla states $\ket{E_{xy}}$ have a specific form in order to fulfill the requirement in Eq.~\eqref{eqn:bell}. They can be written in terms of an orthonormal basis of Eve $\{\ket{f_{i,j}}_E\}$:
\begin{equation}
\ket{E_{xy}}=\begin{cases}\frac{1}{\sqrt{\beta_0}}\sum_{k=0}^{d-1}\sqrt{\lambda_{0,k}}\omega^{xk}\ket{f_{0,k}}_E&\text{for}\; x=y\\
\frac{1}{\sqrt{d}}\sum_{k=0}^{d-1}\omega^{xk}\ket{f_{y-x,k}}_E&\text{for}\; x\neq y, \end{cases}
\end{equation}
with $\omega:=e^{2\pi i/d}$ and $\lambda_{j,k}$ given above.
The ancilla states with $x=y$ have a fixed angle between each other, they are called pyramid states \cite{Lia03}. They fulfill 
\begin{equation*}
\overlap{E_{xy}}{E_{x'y'}}=
\begin{cases} 1 &\text{if}\; x=x'\; \text{and}\; y=y',\\
1-\frac{\beta_1}{\beta_0} &\text{if}\; x=y\neq x'=y',\\
0 &\text{otherwise}. \end{cases}
\end{equation*}

The eavesdropper would like to know Alice's and Bob's classical value $x$ and $y$, respectively. For the case $x\neq y$ she knows both values with certainty, as those ancilla states are orthogonal and she can perfectly discriminate them. For the case $x=y$, Eve has to discriminate $d$ pyramid states. Measurements for such symmetrical states exist, and it is known that the error-minimizing measurement for such states is the \textit{square-root measurement} \cite{Hol78,Hau94,Ban97,Eld01,Che00,Bar01}. The following results for the tomographic protocol were derived in \cite{Lia03,Kas04}:

The state in Eq.~\eqref{eqn:rhoE} can be rewritten as:
\begin{equation}
\rho_{E}=\beta_0 \rho^{(=)}+(1-\beta_0)\rho^{(\neq)},
\label{eq:evetotal}
\end{equation}
where the density operator $\rho^{(=)}=\frac{1}{d}\sum_{x=0}^{d-1} \ket{E_{xx}}\bra{E_{xx}}$ denotes the cases, when Alice and Bob have the same values, whereas in the case of $\rho^{(\neq)}=\frac{1}{d(d-1)}\sum_{y \neq x} \ket{E_{xy}}\bra{E_{xy}}$ their values are different. The eavesdropper wants to find their common values, so she wants to discriminate those ancilla states for $x=y$. The POVM elements $\proj{e_{xx}}$, that discriminate the pyramid states with minimum error, are given via
\begin{equation*}
\ket{e_{xx}}=\frac{1}{\sqrt{d\rho^{(=)}}}\ket{E_{xx}}, 
\end{equation*}
i.e.\ the name \textit{square-root measurement} is related to the construction of the elements. 
An explicit calculation for the operator $\frac{1}{\sqrt{d\rho^{(=)}}}$ results in \cite{Lia03}:
\begin{equation*}
\frac{1}{\sqrt{d\rho^{(=)}}}=\frac{(r_0+\sqrt{r_0 r_1}+r_1)\mathds{1}-\rho^{(=)}}{\sqrt{r_0 r_1}(\sqrt{r_0}+\sqrt{r_1})},
\end{equation*}
where $r_0=1-\frac{d-1}{d}\frac{\beta_1}{\beta_0}$ is the eigenvalue corresponding to the eigenvector $\sum_x \ket{E_{xx}}$ and $r_1=\frac{\beta_1}{d\beta_0}$ is the $(d-1)$-fold eigenvalue for the eigenvector $(\ket{E_{xx}}-\frac{1}{d}\sum_y \ket{E_{yy}})$. From this the overlap $\overlap{e_{xx}}{E_{yy}}$ can be calculated as 
\begin{equation*}
\overlap{e_{xx}}{E_{yy}}=\sqrt{\eta_0}\delta_{xy}+\sqrt{\eta_1}(1-\delta_{xy}),
\end{equation*}
with $\sqrt{\eta_0}=\frac{\sqrt{r_0}+(d-1)\sqrt{r_1}}{\sqrt{d}}$, $\sqrt{\eta_1}=\frac{\sqrt{r_0}+\sqrt{r_1}}{\sqrt{d}}$ and $\delta_{xy}$ the Kronecker delta. The probability $\eta_0$ denotes the probability that Eve, when finding $\ket{e_{xx}}$, knows that Alice and Bob share the value $x$ and $\eta_1$ denotes the probability that they hold one of the other $d-1$ values.

The eavesdropper's probability to guess the right value of Alice consists of the following parts: the probability $(1-\beta_0)$ that the density operator $\rho^{(\neq)}$ appears (see Eq.~\eqref{eq:evetotal}) and the probability $\beta_0$ that $\rho^{(=)}$ appears multiplied with the probability that she guesses the right value in this case, which was $\eta_0$ (see above).
Inserting $r_0$ and $r_1$ into $\eta_0$, we get an expression for the guessing probability depending on $d$ and the error rate $Q=1-\beta_0$:
\begin{widetext}
\begin{equation}
p_\mathrm{guess}^{\textrm{six-state}}(d,Q)=1-\beta_0+\beta_0\eta_0=Q+\frac{(1-Q)}{d} \left[ 1-\frac{(d-2) Q}{d (Q-1)}+2 (d-1) \sqrt{\frac{dQ-(d+1)Q^2}{(d-1)d^2(1-Q)^2}}\right].
\label{eq:pguessd}
\end{equation}
\end{widetext}

\subsection{BB84 for qubit-systems}
A strategy to distinguish two nonorthogonal quantum states is called minimal-error discrimination (MED) (see \cite{Hol73,Yuen75,Hel76,Ban97,Her04}).
%
In MED for each measurement one has a conclusive result, but with probability $p_{\mathrm{err}}$ the result is erroneous. It was shown by Helstr\o{}m \cite{Hel76} that the maximal probability to make a correct guess when distinguishing two quantum states $\rho_E^0$ and $\rho_E^1$ that appear with the same probability $p_0=p_1=\frac{1}{2}$ is given by
\begin{equation}
p_{\mathrm{guess}}(2,Q)=1-p^{\min}_{\mathrm{err}}=\frac{1}{2}\left(1+\frac{1}{2}\norm{ \rho_E^0-\rho_E^1}\right). 
\label{min}
\end{equation}
In order to calculate $\norm{ \rho_E^0-\rho_E^1}$, we express the states $\rho_E^0$ and $\rho_E^1$ (see Eq.~\eqref{eqn:rhoEx}) in terms of the computational basis of Eve.

Assuming that Eve has a purifying system of the state in Eq.~\eqref{eqn:bell2}, and that Alice and Bob perform a von Neumann measurement, one can derive an expression for $\norm{ \rho_E^0-\rho_E^1}$ for the BB84-protocol. The operator $\rho_E^0-\rho_E^1$ can be written as 
\begin{eqnarray}
\rho_E^0-\rho_E^1&=&
2\sqrt{\lambda_{00}\lambda_{01}}\left(\ketbra{00}{01}+\ketbra{01}{00}\right)\nonumber \\
&& +2\sqrt{\lambda_{10}\lambda_{11}}\left(\ketbra{10}{11}+\ketbra{11}{01}\right),
\end{eqnarray}
so 
\begin{eqnarray}
|\rho_E^0-\rho_E^1|&=&2\sqrt{\lambda_{00}\lambda_{01}}\left(\p{00}+\p{01} \right)\nonumber \\
&&+2\sqrt{\lambda_{10}\lambda_{11}}\left(\p{10}+\p{11} \right),
\end{eqnarray}
with $|A|=\sqrt{A^{\dagger}A}$. The eigenvalues 
$2\sqrt{\lambda_{00}\lambda_{01}}$ and $2\sqrt{\lambda_{10}\lambda_{11}}$ occur with multiplicity 2. Thus the 1-norm is 
\begin{equation}
\frac{1}{2}\norm{\rho_E^0-\rho_E^1}=2\sqrt{\lambda_{00}\lambda_{01}}+2\sqrt{\lambda_{10}\lambda_{11}}.
\label{norm:result}
\end{equation}

The error rates in the $z$- and $x$-direction are $e_z=\lambda_{10} +\lambda_{11} $ and $e_x=\lambda_{01} +\lambda_{11}$ (see \cite{RenPhD,Sca09}). There remains one free parameter, that we have to optimize to obtain the best case for Eve. We adopt the method in \cite[Appendix A]{Sca09} to maximize the probability of correct guess in Eq.~(\ref{min}): according to \cite{Sca09}, we choose $\lambda_{00}=(1-Q)(1-u)$, $\lambda_{01}=(1-Q)u$, $\lambda_{10}=Q(1-v)$ and $\lambda_{11}=Q v$, with  $u,v \in [0,1]$ and the additional constraint (from $\lambda_{01}+\lambda_{11}=Q$)
\begin{equation}
(1-Q)u+Q v=Q.
\label{eq:constraint}
\end{equation}
Defining $\ket{\Phi_{ij}}$ as the corresponding Bell states to the value $\lambda_{ij}$, the purification of the state $\rho_{AB}$ can be written as $\ket{\psi_{ABE}}=\sum_{ij} \sqrt{\lambda_{ij}}\ket{\Phi_{ij}}_{AB} \otimes \ket{e_{ij}}_E$, where $\{\ket{e_{ij}}\}$ is a four-dimensional orthonormal basis. Using Eq.~\eqref{norm:result} and the constraint given in Eq.~\eqref{eq:constraint} we find a function, which depends on the parameter $v$:
\begin{eqnarray}
\frac{1}{2}\norm{\rho_E^0- \rho_E^1}&=&f(v):=2\sqrt{ \left(1-v\right)\,Q\,\left[ 1 +\left(v-2\right)Q\right] } \nonumber \\
&&+2\sqrt{\left( 1 -v \right) \,v\,Q^2}. 
\end{eqnarray}

 Finding the maximum of the expression leads to the result $u=v=Q$ and finally to the expressions of $\lambda_{ij}$:
$\lambda_{00}=(1-Q)^2$, $\lambda_{01}=\lambda_{10}=(1-Q)Q$ and $\lambda_{11}=Q^2$.
This gives the guessing probability:
\begin{equation}
p^{\mathrm{BB84}}_{\mathrm{guess}}(2,Q)=\frac{1}{2}\left(1+2\sqrt{(1-Q)Q} \right).
\label{eq:minbb84}
\end{equation}

By using the same methods we can derive the guessing probability for the six-state protocol, which lead to the same result as derived in Eq.~\eqref{eq:pguessd}:
\begin{equation}
p^{\mathrm{six-state}}_{\mathrm{guess}}(2,Q)=\frac{1}{2}\left(1+\sqrt{Q(2-3Q)}+Q \right).
\label{eq:minsix}
\end{equation}

\subsection{Optimal multistate MED measurement from additivity of min-entropy}
We know from \cite{RenPhD} that the min-entropy is additive, i.e.\ for tensor product states $\rho_{XE}^{\otimes n}$ it holds that $H_{\min}(\rho_{XE}^{\otimes n}|E^n)=n H_{\min}(\rho_{XE}|E)$. The min-entropy is a function of the probability of a correct guess of Eve's states. The state $\rho_{XE}^{\otimes n}$ is of the form:

\begin{eqnarray}
\rho_{XE}^{\otimes n}&=&\left(\frac{1}{d}\sum_{x=0}^{d-1}  \proj{x} \otimes  \rho_E^x\right)^{\otimes n}\\
&=& \frac{1}{d^n}\sum_{\textbf{x}\in \{0,...,d-1\}^n} \proj{\textbf{x}} \otimes  \rho_{E^n}^\textbf{x},
\end{eqnarray}
where 
\begin{equation}
\rho_{E^n}^\textbf{x}=\bigotimes_{i=0}^{n-1} \rho_E^{x_i}
\label{eq:productstates}
\end{equation}
and $\textbf{x}=(x_0,...,x_{n-1})$ is a vector of length $n$ with $x_i \in \{0,...,d-1\}$. Thus, Eve's state is given by

\begin{equation}
\rho_E^{\otimes n}=\frac{1}{d^n}\sum_{\textbf{x}}\rho_{E^n}^\textbf{x}
\label{eqn:Even} 
\end{equation}
and is a sum of tensor product states, see Eq.~\eqref{eq:productstates}. The explicit minimum-error discrimination problem is to distinguish the set of states $\{\rho_{E^n}^\textbf{x}\}$ for different $\textbf{x}$. We can conclude from the additivity of the min-entropy, that for the set of states given in Eq.~\eqref{eq:productstates} the optimal MED measurement consists of optimal MED measurements on the single-signal states $\rho_E^{x_i}$. This result is interesting, as in general measurements in the total Hilbert space may lead to higher guessing probabilities than measurements in individual subspaces. To the best of our knowledge, this result is not known in the context of state discrimination.
\section{Comparison of key rates}
\label{sec:comparison}

In this chapter we provide the results of parameter estimation with CPOVM (see Sec.~\ref{sec:parameter}) and those of the calculation of the min-entropy (see Sec.~\ref{sec:guessing}). We first review some results about finite-key distribution.

For a finite number of signals, the achievable secure key rate is found to be \cite{Sca08a,Cai09}:
\begin{equation}
\ell_1/N= \frac{n}{N}\left( S_{\xi}(\rho_{XE}|E)+\Delta-\mathrm{leak}_{\mathrm{EC}} \right)+\frac{2}{N}\log_2\left(2\eps{PA}\right),
\label{r2}
\end{equation}
with $\Delta:=-7\sqrt{\frac{\log_2{\left(2/\bar{\varepsilon}\right)}}{n}}$, the total security parameter $\eps{}$ (see e.g.\ \cite{Sca08,Cai09}) 
\begin{equation}
\eps{}=\eps{PA}+\eps{EC}+\eps{PE}+\epsbar
\label{totalsecurity}
\end{equation} 
and $S_{\xi}(\rho_{XE}|E):=\min_{\bar{\rho}_{XE}\in \Gamma_{\xi}}S(\rho_{XE}|E) $. The set $\Gamma_{\xi}=\{\sigma: \frac{1}{2}|\lambda_m -\lambda_{\infty} (\sigma)|\leq \xi\}$ contains all states compatible with the statistics in parameter estimation. The conditional von Neumann entropy with the correction term $\Delta$ is a lower bound on the $\epsbar$-smooth min-entropy. The leakage term $\mathrm{leak}_{\mathrm{EC}}$ is taken from \cite{Sca09} to be $\mathrm{leak}_{\mathrm{EC}}=1.2 h(Q)$ for $\eps{EC}=10^{-10}$, where $h(x)$ is the binary entropy. Throughout all calculations, we assume asymmetric protocols with a symmetric attack. An asymmetric protocol means that one only keeps the measurement results of one particular basis for the key; the other results are used for parameter estimation. In the case of protocols with $(d+1)$ bases (e.g.\ the six-state protocol with $d=2$) this basis is chosen with probability $q=(1-d p)$ and the other $d$ bases with probability $p$. For protocols with 2 bases (e.g.\ the BB84-protocol with $d=2$) $q=1-p$. Taking the largest deviation $\xi_i$ from the perfect parameter in one measurement basis and equating it with the other deviations, leads to a symmetric choice of parameters $m_i$ and $\eps{PE_i}$, i.e.\ $m_i=\frac{m}{d+1}$ ($m_i=\frac{m}{2}$ for $2-$bases protocols) and $\eps{PE_i}=\frac{\eps{PE}}{d+1}$ ($\eps{PE_i}=\frac{\eps{PE}}{2}$) (see section~\ref{sec:parameter}). This assumption gives a lower bound on the secret key rate. The number of signals used for parameter estimation is given by $m=Np^2$. 

In order to calculate the key rate, we fix $\eps{}$ and $\eps{EC}$ and  maximize $\ell_1/N$ in Eq.~\eqref{r2} for the parameters $\eps{P
E},\eps{PA},\epsbar$ and $q$ with a computational software program (Mathematica) under the constraint given in Eq.~\eqref{totalsecurity}.


\subsection{Key rates via von Neumann entropy for different approaches of parameter estimation}

For a comparison of the approaches (IPOVM, CPOVM) explained in Sec.~\ref{sec:parameter} we consider the asymmetric BB84- and six-state-protocol for a symmetric attack for dimension $d=2$ as discussed in \cite{Sca08a}. 
In the calculation of the key rates via the von Neumann entropy (see Eq.~\eqref{r2}) we use a QBER of $Q=0.05$ and a total security parameter of $\eps{}=10^{-9}$ (see Eq.~\eqref{totalsecurity}). 
The conditional von Neumann entropy for the six-state protocol is given by \cite{Sca08a,Sca09}:
\begin{equation}
S^{\mathrm{six-state}}(\rho_{XE}|E)=\left(1-Q\right)\left[1-h\left(\frac{1-\frac{3}{2}Q}{1-Q} \right) \right]
\label{vNeumannsix} 
\end{equation}
and for the BB84-protocol
\begin{equation}
S^{\mathrm{BB84}}(\rho_{XE}|E)=1-h(Q).
\label{vNeumannBB84} 
\end{equation}

The variables $\xi$ for parameter estimation used in this comparison are summarized in Tab.~\ref{tab:PE}. Note that the symmetrized state is parametrized by only one parameter. This has no influence on the IPOVM approach, in contrast to CPOVM, where the number of POVM outcomes can be reduced from 3 for BB84 (4 for six-state) to 2 (2). 

\begin{table}[h]
\begin{ruledtabular}
\begin{tabular}{c|cc}
 \nonumber & BB84 & six-state \\ \hline
 IPOVM & $\xi(\frac{\eps{PE}}{2},2,\frac{m}{2})$ & $\xi(\frac{\eps{PE}}{3},2,\frac{m}{3})$ \\ 
 CPOVM & $\xi(\eps{PE},2,m)$ & $\xi(\eps{PE},2,m)$ \\ 
\end{tabular}
\caption{\label{tab:PE} Deviations $\xi$ from perfect parameter (see Eq.~\eqref{PE} in Sec.~\ref{sec:parameter}) for different parameter estimation approaches (IPOVM and CPOVM): BB84 and six-state protocol.}
\end{ruledtabular}
\end{table}

\begin{figure}[h]
  \centering
  \includegraphics[width=8cm]{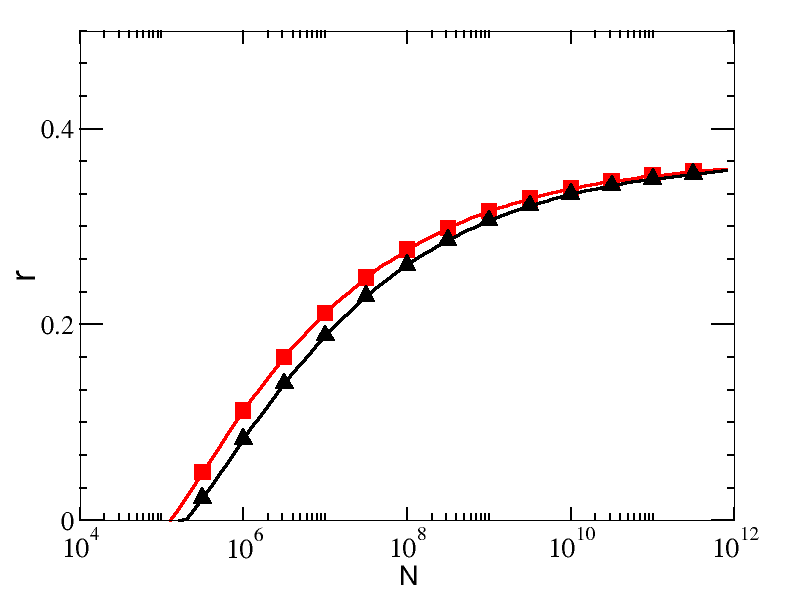}
  \caption{(Color online) Comparison of the key rates (calculated via the von Neumann entropy, see Eqs.~\eqref{r2} and \eqref{vNeumannBB84}) using different parameter estimations for asymmetric BB84-protocol; $\eps{}=10^{-9}$, $Q=5\text{\%}$; squares (red): CPOVM, triangles (black): IPOVM (see Sec.~\ref{sec:parameter} for explanations).}
  \label{fig:PEplot1}
 \end{figure}
\begin{figure}[h]
  \centering
  \includegraphics[width=8cm]{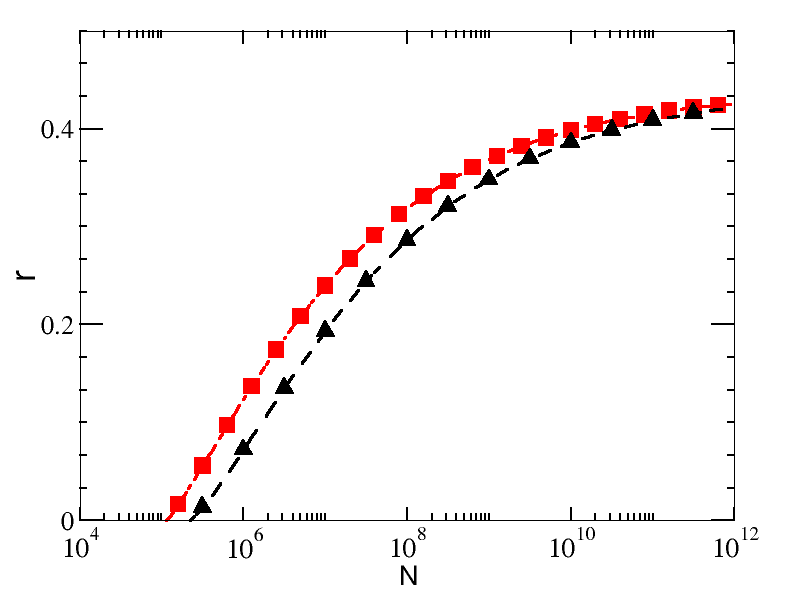}
  \caption{(Color online) Comparison of the key rates (calculated via the von Neumann entropy, see Eqs.~\eqref{r2} and \eqref{vNeumannsix}) using different parameter estimations for asymmetric six-state-protocol; $\eps{}=10^{-9}, Q=5\text{\%}$; squares (red): CPOVM, triangles (black): IPOVM (see Sec.~\ref{sec:parameter} for explanations).}
  \label{fig:PEplot2}
 \end{figure}

The results are shown in Fig.~\ref{fig:PEplot1} and ~\ref{fig:PEplot2}. We point out that our CPOVM approach leads to higher key rates for the BB84- and six-state-protocol. In particular for signals $N \lesssim 10^{11}$, the numerical analysis reveals the importance of parameter estimation. While the CPOVM approach leads for $N=10^6$ signals to a $72\%$ (35\%) higher key rate than the IPOVM approach for the six-state- (BB84-) protocol, the improvement for $N=10^{10}$ is still $3 \%$ (2\%).

%

\subsection{Key rates via the min-entropy for two-dimensional quantum systems}
In this section we exploit the preceding results from Sec.~\ref{sec:guessing} regarding the min-entropy in order to compute the secret key rate and compare it to the key rate calculated with Eq.~\eqref{r2}.

We explained in Sec.~\ref{sec:privacy} that the achievable upper bound on the secure key length $\ell$ after the \textit{privacy amplification} step is given by Eq.~\eqref{eq:keylength}.
We can derive a key rate by using the following bounds \cite[Lemma 3.2.6]{RenPhD}: 
\begin{equation}
H^{\epsbar}_{\min}(\rho_{XE}^{\otimes n}|E^n)\geq n H_{\min}^{\epsbar/n}(\rho_{XE}|E)\geq nH_{\min}(\rho_{XE}|E);
\label{bound} 
\end{equation}
the last inequality is a very good approximation as $\epsbar$ is in the order of $10^{-10}$. Thus, we arrive at the following key rate:
\begin{equation}
\ell_2/N= \frac{n}{N}\left(H_{\min,\xi}(\rho_{XE}|E)-\mathrm{leak}_{\mathrm{EC}}\right) +\frac{2}{N}\log_2\left(2\eps{PA}\right),
\label{r1}
\end{equation}
where the leakage term $\mathrm{leak}_{\mathrm{EC}}$ and $\eps{PA}$ are the same as in Eq.~\eqref{r2}, and $H_{\min,\xi}(\rho_{XE}|E):=\min_{\bar{\rho}_{XE}\in \Gamma_{\xi}}H_{\min}(\rho_{XE}|E)$ (see Eq.~\eqref{r2}). We calculate this key rate using the connection to the guessing probability, i.e.\ $H_{\min}^{\mathrm{protocol}}(\rho_{XE}|E)=-\log_2 \text{p}_\text{guess}^\mathrm{protocol}$ (see Eq.~\eqref{eq:min-guess}), and compare it to the key rate given in Eq.~\eqref{r2}. The guessing probability for the specific protocol is given by Eqs.~\eqref{eq:minbb84} and \eqref{eq:minsix}.

\begin{figure}[h]
 \centering
 \includegraphics[width=8cm]{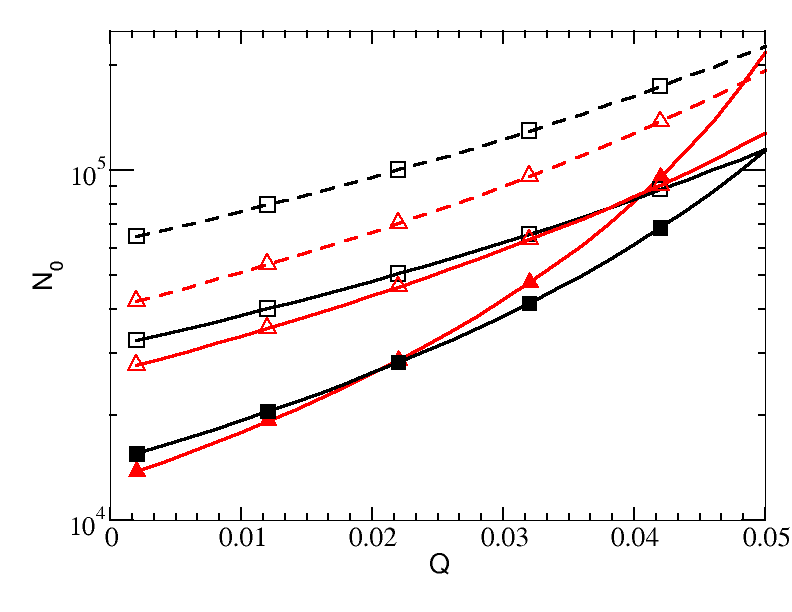}
\caption{(Color online) Threshold value $N_0$ (number of signals, where the key rate becomes non-zero) vs QBER $Q$ with $\varepsilon=10^{-9}$ and $\eps{EC}=10^{-10}$; triangles (red): BB84-protocol, squares (black): six-state-protocol; filled: min-entropy (Eq.~\eqref{r1}), open: von Neumann entropy with CPOVM approach (Eq.~\eqref{r2}), dashed line: von Neumann entropy (Eq.~\eqref{r2}) with IPOVM approach (see Sec.~\ref{sec:parameter} for explanations).}
\label{fig:plot}
\end{figure}

In Fig.~\ref{fig:plot} the threshold number of signals $N_0$, where the key rate becomes non-zero, is plotted as a function of the QBER $Q$. For parameter estimation, we have considered the CPOVM approach (see Sec.~\ref{sec:parameter}) with the variables given in Tab.~\ref{tab:PE}. Additionally, we have plotted the key rate via the von Neumann entropy (Eq.~\eqref{r2}) for the IPOVM approach.
In comparison to the von Neumann approximation (Eq.~\eqref{r2}), only 1/2 (7/10) of the number of signals is needed for non-zero key rates in the six-state-protocol for $Q=0.2\%$ ($Q=3.8\%$), when using the min-entropy. For the BB84-protocol, only 1/2 (9/10)  of the number of signals is needed for $Q=0.2\%$ ($Q=3.8\%$).

Thus, by calculating a key rate explicitly with the min-entropy, we get positive key rates for a smaller number of signals than via the von Neumann entropy approach. This behavior can be explained by the correction term $\Delta$ in the key rate in Eq.~\eqref{r2}. For a small number of total signals $N$, this correction term is not a good approximation and has a big impact on the key rate.

We point out that for low $Q$ we can achieve non-zero key rates with only $\mathcal O(10^4)-\mathcal O(10^5)$ signals. Note that in \cite{Scarani2010} it was considered a ``milestone" to reach non-zero key rates for significantly less than $10^5-10^6$ signals.

\subsection{Key rates via the min-entropy for $d$-dimensional quantum systems}
In \cite{She10} the influence of the dimension on the key rate was discussed. Exploiting the results from this paper, we discuss the improvement for higher-dimensional quantum systems. Throughout this part we only consider the $(d+1)$-bases protocols, such as the six-state-protocol for $d=2$. 
Furthermore, we adapt our CPOVM approach and by using Eq.~\eqref{PE} from Sec.~\ref{sec:parameter} we get $\xi\left(\eps{PE},2,m\right)$.
The correction term to the $d$-dimensional von Neumann entropy is given in \cite{Sheridan2011} as 
$\Delta=-(2 \log_2 d+3)\sqrt{\frac{\log_2 (2/\epsbar)}{n}}$ and the leakage term is characterized by $\mathrm{leak_{EC}}=1.2h_d(Q)$ with $h_d(p):=-p \log_2 \left(\frac{p}{d-1}\right)-(1-p)\log_2 (1-p)$. The conditional von Neumann entropy was calculated in \cite{She10} as 
\begin{equation}
 S^d(\rho_{XE}|E)= (1-Q)\left[\log_2 d-h_d\left(1-\frac{1-\frac{d+1}{d}Q}{1-Q}\right)\right],
\label{eqn:dvNeumann}
\end{equation}
where $Q=1-\beta_0$ denotes the error rate in the sifted key. We will compare the key rate calculated via the $d$-dimensional conditional von Neumann entropy, with the one via the $d$-dimensional min-entropy. The latter can be obtained by using
\begin{equation}
H_{\min}^d(\rho_{XE}|E)=-\log_2 p_\mathrm{guess}(d,Q), 
\label{eqn:dmin}
\end{equation}
where $p_\mathrm{guess}(d,Q)$ was given in Eq.~\eqref{eq:pguessd}. 

In order to quantify the number of signals, we have scaled $N_0$ with $\log_2 d$, as for example sending one state in the dimension $d=4$ corresponds to sending two states in the dimension $d=2$. For making the key rate comparable to the two-dimensional case, it has to be divided by $\log_2 d$. The dimensions are prime numbers as complete mutually unbiased bases can be formed for primes and prime powers (see e.g.\ review article \cite{Dur10}).

\begin{figure}[h]
 \centering
 \includegraphics[width=8cm]{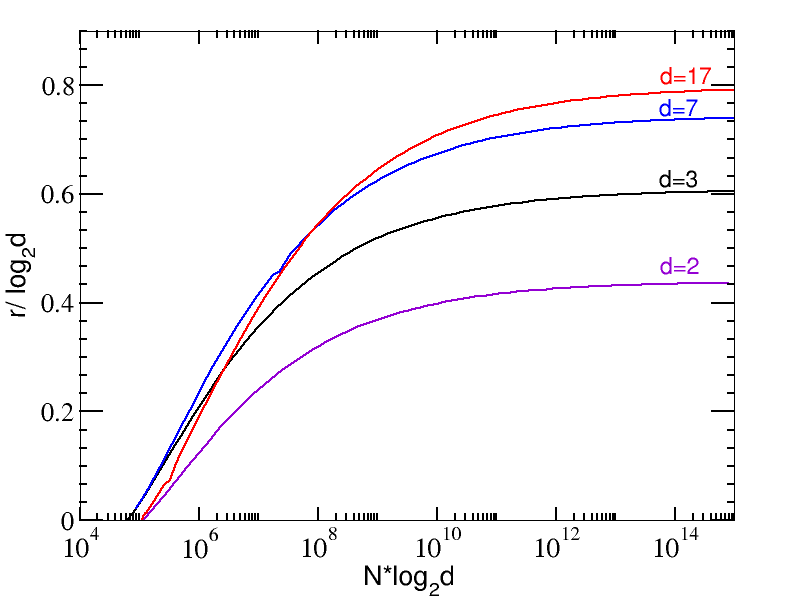}
\caption{(Color online) Key rates with $d$-dimensional conditional von Neumann entropy (Eq.~\eqref{eqn:dvNeumann}) plotted versus scaled total number of signals for a fixed error rate $Q=5\%$. This is analogous to \cite{She10}, where a different scale was used for the axes.}
\label{fig:dvonNeumann}
\end{figure}

\begin{figure}[h]
 \centering
 \includegraphics[width=8cm]{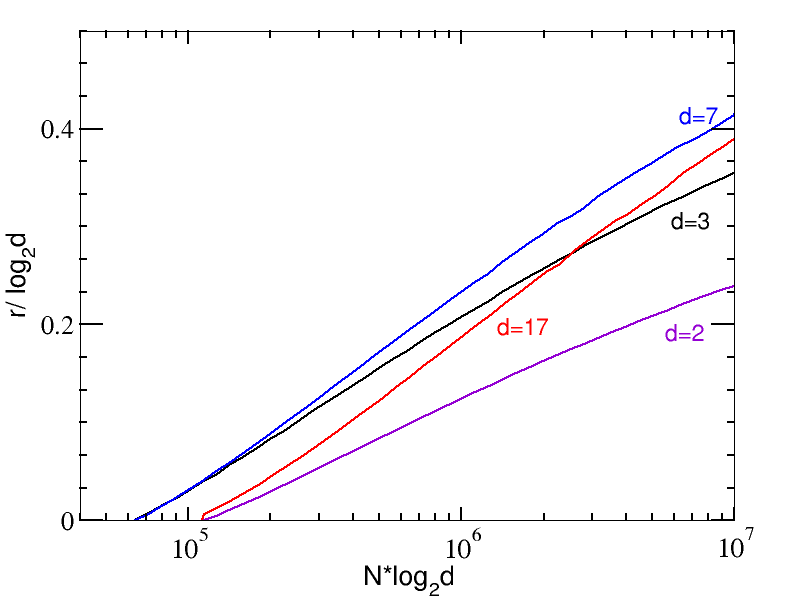}
\caption{(Color online) Key rates with $d$-dimensional conditional von Neumann entropy (Eq.~\eqref{eqn:dvNeumann}) plotted versus scaled total number of signals for a fixed error rate $Q=5\%$ (Magnification of Fig.~\ref{fig:dvonNeumann}).}
\label{fig:dvonNeumannG}
\end{figure}

Figure~\ref{fig:dvonNeumann} shows the behavior of the key rate calculated with Eq.~\eqref{eqn:dvNeumann} for different dimensions. In contrast to \cite{She10}, we scaled the key rate with the dimension. It can be seen from the plot, that higher dimensions are advantageous as the key rate increases. In order to obtain the behavior for a small number of signals, Fig.~\ref{fig:dvonNeumannG} provides a magnification of this area. The higher the dimension, the more the point, where the key rate becomes non-zero is shifted to the right (apart from the case $d=2$). This might be due to the correction term, as it scales linearly with the dimension, so for higher dimension, more is subtracted from the conditional von Neumann entropy. We will see in the next paragraph, that the min-entropy approach has an advantage over the von Neumann  entropy approach for a small number of signals.

\begin{figure}[h]
 \centering
 \includegraphics[width=8cm]{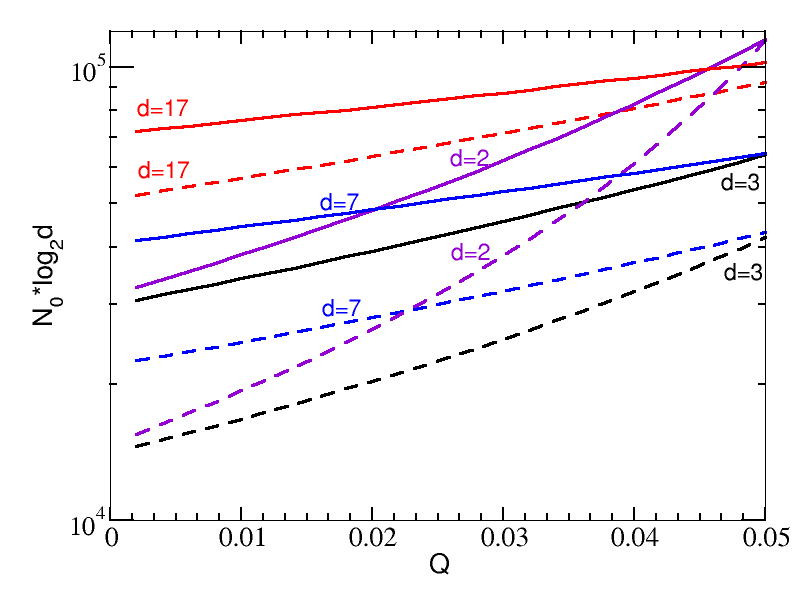}
\caption{(Color online) Threshold value $N_0$ (number of signals, where the key rate is positive) vs QBER with $\varepsilon=10^{-9}$ and $\eps{EC}=10^{-10}$ for different dimensions $d \in \{2,3,7,17\}$. Dashed line: min-entropy (Eq.~\eqref{eqn:dmin}), straight line: von Neumann entropy (Eq.~\eqref{eqn:dvNeumann}).}
\label{fig:plotd}
\end{figure}

In Fig.~\ref{fig:plotd} we compare the number $N_0$, where the key rate becomes non-zero, for key rates using the quantities given in Eqs.~\eqref{eqn:dvNeumann} and \eqref{eqn:dmin} for different dimensions. It can be seen that the min-entropy approach is better throughout the presented error rates. The advantage of the min-entropy approach (Eq.~\eqref{eqn:dmin})  over the von Neumann approach (Eq.~\eqref{eqn:dvNeumann}) augments with increasing dimensions. This can be explained again with the correction term that scales linearly with the dimension. When comparing higher dimensions to the qubit case, one can see that for certain error rates the dimensions bigger than two are advantegous. The dimension $d=3$ for example, gives a lower threshold value $N_0$ for non-zero key rates than the qubit case throughout all the presented error rates.

\section{Conclusion}
\label{sec:conclusion}

We have improved the secret key rates in QKD with a finite number of signals, by considering parameter estimation to be implemented by a single POVM for all parameters. Additionally, we have calculated the min-entropy for a single-signal state in $d$-dimensions explicitly by using its operational meaning via the guessing probability. We showed that using this ansatz for a small number of signals leads to computable non-zero key rates. This advantage of the min-entropy might be due to the correction term $\Delta$ in key rate calculations using the conditional von Neumann entropy \cite{Sca08a,Cai09,She10}, as this correction is big for a small number of signals. This correction term scales linearly with the dimension, so an improvement for high dimensions (up to $d=17$) is found by calculating the min-entropy. Thus higher-dimensional systems might be advantageous when resources are limited. As a spin-off, we have deduced from the additivity property of the min-entropy and its relation to the guessing probability, that the optimal minimum-error discrimination measurement (MED) for a set of tensor product states with a certain symmetry is the optimal MED measurement on each subsystem.

Considering the importance of finite-key analysis for practical implementations, we have shown that non-zero secure key rates can be achieved already with $10^4-10^5$ signals per run.

\begin{acknowledgments}
We would like to thank Silvio Abruzzo, Fabian Furrer, Matthias Kleinmann, and in particular Renato Renner for valuable and enlightening discussions. This work was financially supported in part by Deutsche Forschungsgemeinschaft (DFG).
\end{acknowledgments}

\begin{appendix}
\section{Proof of Theorem~\ref{thm:1}}
\label{appendix:proof}

\begin{proof}
We first show that $\Prob{\dist{\vec{\lambda}_m-\vec{\lambda}_\infty}>2\xi}\leq \eps{PE}$. Starting from the law of large numbers \cite{Cov91},
 \begin{equation}\label{largenumbers}
  \Prob{D(\vec{\lambda}_m||\vec{\lambda}_\infty)>2\xi'}\leq 2^{-m\left(2\xi'-|\chi|\frac{\log_2{(m+1)}}{m}\right)},
 \end{equation}
with $D(\vec{\lambda}_m||\vec{\lambda}_\infty):=\sum_{i=1}^{|\chi|}\lambda_m(i)\log_2\left(\frac{\lambda_m(i)}{\lambda_\infty(i)} \right)$
 and using \cite{Cov91}
 \begin{equation}\label{reldist} 
  \dist{\vec{\lambda}_m-\vec{\lambda}_\infty}\leq \sqrt{\frac{D(\vec{\lambda}_m||\vec{\lambda}_\infty)\ln{2}}{2}},
 \end{equation}
 we result in
 \begin{eqnarray}
  &\nonumber&\Prob{\dist{\vec{\lambda}_m-\vec{\lambda}_\infty}>\sqrt{\frac{2\xi'\ln{2}}{2}}} \nonumber \\ 
  &\stackrel{(\ref{reldist})}{\leq}& \Prob{\sqrt{\frac{D(\vec{\lambda}_m||\vec{\lambda}_\infty)\ln{2}}{2}}>\sqrt{\frac{2\xi'\ln{2}}{2}}} \nonumber \\
  &\stackrel{(\ref{largenumbers})}{\leq}& \label{PE1} 2^{-m\left(2\xi'-|\chi|\frac{\log{(m+1)}}{m}\right)}.
 \end{eqnarray}
 For $\xi:=\sqrt{\frac{2\xi'\ln{2}}{2}}$ it follows:
  \begin{eqnarray*}
   &\nonumber&\Prob{\dist{\vec{\lambda}_m-\vec{\lambda}_\infty}>2\xi} \\
   &=& \Prob{\dist{\vec{\lambda}_m-\vec{\lambda}_\infty}>\sqrt{\frac{2 (4\xi')\ln{2}}{2}}} \\
   &\stackrel{(\ref{PE1})}{\leq}& 2^{-m\left(2 (4\xi')-|\chi|\frac{\log{(m+1)}}{m}\right)} \\
   &=&  2^{-m \left(8\frac{\xi^2}{\ln{2}}-|\chi|\frac{\log{(m+1)}}{m}\right)}=: \eps{PE}.
  \end{eqnarray*}
Then except with probability $\eps{PE}$, the following holds:
 \begin{equation*}
  \dist{\vec{\lambda}_m-\vec{\lambda}_\infty}\leq 2 \xi 
 \end{equation*}
 with $\xi=\sqrt{\frac{\ln{\left(\frac{1}{\eps{PE}}\right)}+|\chi|\ln{(m+1)}}{8m}}$.
 It remains to show that 
 \begin{equation*}
  \dist{\lambda_m-\lambda_\infty}\equiv\frac{1}{2}|\lambda_m-\lambda_\infty|\leq \frac{1}{2}\dist{\vec{\lambda}_m-\vec{\lambda}_\infty}.
 \end{equation*} 
 Remember that we denote by $\lambda_m:=\lambda_m(k)$ and $\lambda_\infty:=\lambda_\infty(k)$ any $k$-th parameter. The normalization conditions of the POVM $\sum_{i=1}^{|\chi|}\lambda_\infty(i)=1=\sum_{i=1}^{|\chi|}\lambda_m(i)$ lead to
 \begin{eqnarray}
  |\lambda_m-\lambda_\infty|
  &=&\left|\sum_{i=1,i\ne k}^{|\chi|}\lambda_m(i)-\lambda_\infty(i)\right| \nonumber \\
  &\stackrel{\Delta}{\leq}&\label{norm}\sum_{i=1,i\ne k}^{|\chi|}\left|\lambda_m(i)-\lambda_\infty(i)\right|
 \end{eqnarray}
 and
 \begin{equation*}
  \sum_{i=1}^{|\chi|}|\lambda_m(i)-\lambda_\infty(i)|\stackrel{(\ref{norm})}{\geq} 2|\lambda_m-\lambda_\infty|.
 \end{equation*}
 The assertion follows by multiplication with factor $\frac{1}{4}$.
\end{proof}
\end{appendix}

\bibliography{Verzeichnis_paper}

\begin{thebibliography}{45}%
\makeatletter
\providecommand \@ifxundefined [1]{%
 \@ifx{#1\undefined}
}%
\providecommand \@ifnum [1]{%
 \ifnum #1\expandafter \@firstoftwo
 \else \expandafter \@secondoftwo
 \fi
}%
\providecommand \@ifx [1]{%
 \ifx #1\expandafter \@firstoftwo
 \else \expandafter \@secondoftwo
 \fi
}%
\providecommand \natexlab [1]{#1}%
\providecommand \enquote  [1]{``#1''}%
\providecommand \bibnamefont  [1]{#1}%
\providecommand \bibfnamefont [1]{#1}%
\providecommand \citenamefont [1]{#1}%
\providecommand \href@noop [0]{\@secondoftwo}%
\providecommand \href [0]{\begingroup \@sanitize@url \@href}%
\providecommand \@href[1]{\@@startlink{#1}\@@href}%
\providecommand \@@href[1]{\endgroup#1\@@endlink}%
\providecommand \@sanitize@url [0]{\catcode `\\12\catcode `\$12\catcode
  `\&12\catcode `\#12\catcode `\^12\catcode `\_12\catcode `\%12\relax}%
\providecommand \@@startlink[1]{}%
\providecommand \@@endlink[0]{}%
\providecommand \url  [0]{\begingroup\@sanitize@url \@url }%
\providecommand \@url [1]{\endgroup\@href {#1}{\urlprefix }}%
\providecommand \urlprefix  [0]{URL }%
\providecommand \Eprint [0]{\href }%
\providecommand \doibase [0]{http://dx.doi.org/}%
\providecommand \selectlanguage [0]{\@gobble}%
\providecommand \bibinfo  [0]{\@secondoftwo}%
\providecommand \bibfield  [0]{\@secondoftwo}%
\providecommand \translation [1]{[#1]}%
\providecommand \BibitemOpen [0]{}%
\providecommand \bibitemStop [0]{}%
\providecommand \bibitemNoStop [0]{.\EOS\space}%
\providecommand \EOS [0]{\spacefactor3000\relax}%
\providecommand \BibitemShut  [1]{\csname bibitem#1\endcsname}%
\let\auto@bib@innerbib\@empty
\bibitem [{\citenamefont {Bennett}\ and\ \citenamefont
  {Brassard}(1984)}]{BB84}%
  \BibitemOpen
  \bibfield  {author} {\bibinfo {author} {\bibfnamefont {C.}~\bibnamefont
  {Bennett}}\ and\ \bibinfo {author} {\bibfnamefont {G.}~\bibnamefont
  {Brassard}},\ }in\ \href@noop {} {\emph {\bibinfo {booktitle} {Proceedings of
  the IEEE International Conference on Computers, Systems, and Signal
  Processing, Bangalore, India}}}\ (\bibinfo  {publisher} {IEEE},\ \bibinfo
  {address} {New York},\ \bibinfo {year} {1984})\ p.\ \bibinfo {pages}
  {175}\BibitemShut {NoStop}%
\bibitem [{\citenamefont {Ben-Or}\ and\ \citenamefont {Mayers}()}]{Ben04}%
  \BibitemOpen
  \bibfield  {author} {\bibinfo {author} {\bibfnamefont {M.}~\bibnamefont
  {Ben-Or}}\ and\ \bibinfo {author} {\bibfnamefont {D.}~\bibnamefont
  {Mayers}},\ }\href@noop {} {\enquote {\bibinfo {title} {General security
  definition and composability for quantum classical protocols},}\ }\Eprint
  {http://arxiv.org/abs/arXiv:quant-ph/0409062} {arXiv:quant-ph/0409062}
  \BibitemShut {NoStop}%
\bibitem [{\citenamefont {Unruh}()}]{Unr04}%
  \BibitemOpen
  \bibfield  {author} {\bibinfo {author} {\bibfnamefont {D.}~\bibnamefont
  {Unruh}},\ }\href@noop {} {\enquote {\bibinfo {title} {Simutable security for
  quantum protocols},}\ }\Eprint {http://arxiv.org/abs/arXiv:quant-ph/0409125}
  {arXiv:quant-ph/0409125} \BibitemShut {NoStop}%
\bibitem [{\citenamefont {M\"uller-Quade}\ and\ \citenamefont
  {Renner}(2009)}]{Mue09}%
  \BibitemOpen
  \bibfield  {author} {\bibinfo {author} {\bibfnamefont {J.}~\bibnamefont
  {M\"uller-Quade}}\ and\ \bibinfo {author} {\bibfnamefont {R.}~\bibnamefont
  {Renner}},\ }\href {\doibase 10.1088/1367-2630/11/8/085006} {\bibfield
  {journal} {\bibinfo  {journal} {New J.\ Phys.}\ }\textbf {\bibinfo {volume}
  {11}},\ \bibinfo {pages} {085006} (\bibinfo {year} {2009})}\BibitemShut
  {NoStop}%
\bibitem [{\citenamefont {Renner}\ and\ \citenamefont
  {K\"{o}nig}(2005)}]{RenKoe}%
  \BibitemOpen
  \bibfield  {author} {\bibinfo {author} {\bibfnamefont {R.}~\bibnamefont
  {Renner}}\ and\ \bibinfo {author} {\bibfnamefont {R.}~\bibnamefont
  {K\"{o}nig}},\ }in\ \href {\doibase 10.1007/978-3-540-30576-7_22} {\emph
  {\bibinfo {booktitle} {Theory of Cryptography}}},\ \bibinfo {series} {Lecture
  Notes in Computer Science}, Vol.\ \bibinfo {volume} {3378},\ \bibinfo
  {editor} {edited by\ \bibinfo {editor} {\bibfnamefont {J.}~\bibnamefont
  {Kilian}}}\ (\bibinfo  {publisher} {Springer Berlin / Heidelberg},\ \bibinfo
  {year} {2005})\ pp.\ \bibinfo {pages} {407--425},\ \Eprint
  {http://arxiv.org/abs/arXiv:quant-ph/0403133} {arXiv:quant-ph/0403133}
  \BibitemShut {NoStop}%
\bibitem [{\citenamefont {Renner}\ \emph {et~al.}(2005)\citenamefont {Renner},
  \citenamefont {Gisin},\ and\ \citenamefont {Kraus}}]{Ren05a}%
  \BibitemOpen
  \bibfield  {author} {\bibinfo {author} {\bibfnamefont {R.}~\bibnamefont
  {Renner}}, \bibinfo {author} {\bibfnamefont {N.}~\bibnamefont {Gisin}}, \
  and\ \bibinfo {author} {\bibfnamefont {B.}~\bibnamefont {Kraus}},\ }\href
  {\doibase 10.1103/PhysRevA.72.012332} {\bibfield  {journal} {\bibinfo
  {journal} {Phys. Rev. A}\ }\textbf {\bibinfo {volume} {72}},\ \bibinfo
  {pages} {012332} (\bibinfo {year} {2005})}\BibitemShut {NoStop}%
\bibitem [{\citenamefont {Kraus}\ \emph {et~al.}(2005)\citenamefont {Kraus},
  \citenamefont {Gisin},\ and\ \citenamefont {Renner}}]{Kra05}%
  \BibitemOpen
  \bibfield  {author} {\bibinfo {author} {\bibfnamefont {B.}~\bibnamefont
  {Kraus}}, \bibinfo {author} {\bibfnamefont {N.}~\bibnamefont {Gisin}}, \ and\
  \bibinfo {author} {\bibfnamefont {R.}~\bibnamefont {Renner}},\ }\href
  {\doibase 10.1103/PhysRevLett.95.080501} {\bibfield  {journal} {\bibinfo
  {journal} {Phys.\ Rev.\ Lett.}\ }\textbf {\bibinfo {volume} {95}},\ \bibinfo
  {pages} {080501} (\bibinfo {year} {2005})}\BibitemShut {NoStop}%
\bibitem [{\citenamefont {Hayashi}(2007)}]{Hay07}%
  \BibitemOpen
  \bibfield  {author} {\bibinfo {author} {\bibfnamefont {M.}~\bibnamefont
  {Hayashi}},\ }\href {\doibase 10.1103/PhysRevA.76.012329} {\bibfield
  {journal} {\bibinfo  {journal} {Phys. Rev. A}\ }\textbf {\bibinfo {volume}
  {76}},\ \bibinfo {pages} {012329} (\bibinfo {year} {2007})}\BibitemShut
  {NoStop}%
\bibitem [{\citenamefont {Scarani}\ and\ \citenamefont
  {Renner}(2008{\natexlab{a}})}]{Sca08a}%
  \BibitemOpen
  \bibfield  {author} {\bibinfo {author} {\bibfnamefont {V.}~\bibnamefont
  {Scarani}}\ and\ \bibinfo {author} {\bibfnamefont {R.}~\bibnamefont
  {Renner}},\ }\href {\doibase 10.1103/PhysRevLett.100.200501} {\bibfield
  {journal} {\bibinfo  {journal} {Phys.\ Rev.\ Lett.}\ }\textbf {\bibinfo
  {volume} {100}},\ \bibinfo {pages} {200501} (\bibinfo {year}
  {2008}{\natexlab{a}})}\BibitemShut {NoStop}%
\bibitem [{\citenamefont {Renner}(2008)}]{RenPhD}%
  \BibitemOpen
  \bibfield  {author} {\bibinfo {author} {\bibfnamefont {R.}~\bibnamefont
  {Renner}},\ }\href {\doibase 10.1142/S0219749908003256} {\bibfield  {journal}
  {\bibinfo  {journal} {Int.\ J.\ Quant.\ Inf.}\ }\textbf {\bibinfo {volume}
  {6}},\ \bibinfo {pages} {1} (\bibinfo {year} {2008})}\BibitemShut {NoStop}%
\bibitem [{\citenamefont {Gisin}\ \emph {et~al.}(2002)\citenamefont {Gisin},
  \citenamefont {Ribordy}, \citenamefont {Tittel},\ and\ \citenamefont
  {Zbinden}}]{Gisin2002}%
  \BibitemOpen
  \bibfield  {author} {\bibinfo {author} {\bibfnamefont {N.}~\bibnamefont
  {Gisin}}, \bibinfo {author} {\bibfnamefont {G.}~\bibnamefont {Ribordy}},
  \bibinfo {author} {\bibfnamefont {W.}~\bibnamefont {Tittel}}, \ and\ \bibinfo
  {author} {\bibfnamefont {H.}~\bibnamefont {Zbinden}},\ }\href {\doibase
  10.1103/RevModPhys.74.145} {\bibfield  {journal} {\bibinfo  {journal} {Rev.
  Mod. Phys.}\ }\textbf {\bibinfo {volume} {74}},\ \bibinfo {pages} {145}
  (\bibinfo {year} {2002})}\BibitemShut {NoStop}%
\bibitem [{\citenamefont {Scarani}\ \emph {et~al.}(2009)\citenamefont
  {Scarani}, \citenamefont {Bechmann-Pasquinucci}, \citenamefont {Cerf},
  \citenamefont {Du\ifmmode~\check{s}\else \v{s}\fi{}ek}, \citenamefont
  {L\"utkenhaus},\ and\ \citenamefont {Peev}}]{Sca09}%
  \BibitemOpen
  \bibfield  {author} {\bibinfo {author} {\bibfnamefont {V.}~\bibnamefont
  {Scarani}}, \bibinfo {author} {\bibfnamefont {H.}~\bibnamefont
  {Bechmann-Pasquinucci}}, \bibinfo {author} {\bibfnamefont {N.}~\bibnamefont
  {Cerf}}, \bibinfo {author} {\bibfnamefont {M.}~\bibnamefont
  {Du\ifmmode~\check{s}\else \v{s}\fi{}ek}}, \bibinfo {author} {\bibfnamefont
  {N.}~\bibnamefont {L\"utkenhaus}}, \ and\ \bibinfo {author} {\bibfnamefont
  {M.}~\bibnamefont {Peev}},\ }\href {\doibase 10.1103/RevModPhys.81.1301}
  {\bibfield  {journal} {\bibinfo  {journal} {Rev. Mod. Phys.}\ }\textbf
  {\bibinfo {volume} {81}},\ \bibinfo {pages} {1301} (\bibinfo {year}
  {2009})}\BibitemShut {NoStop}%
\bibitem [{\citenamefont {Meyer}\ \emph {et~al.}(2006)\citenamefont {Meyer},
  \citenamefont {Kampermann}, \citenamefont {Kleinmann},\ and\ \citenamefont
  {Bru{\ss}}}]{Mey06}%
  \BibitemOpen
  \bibfield  {author} {\bibinfo {author} {\bibfnamefont {T.}~\bibnamefont
  {Meyer}}, \bibinfo {author} {\bibfnamefont {H.}~\bibnamefont {Kampermann}},
  \bibinfo {author} {\bibfnamefont {M.}~\bibnamefont {Kleinmann}}, \ and\
  \bibinfo {author} {\bibfnamefont {D.}~\bibnamefont {Bru{\ss}}},\ }\href
  {\doibase 10.1103/PhysRevA.74.042340} {\bibfield  {journal} {\bibinfo
  {journal} {Phys.\ Rev.\ A}\ }\textbf {\bibinfo {volume} {74}},\ \bibinfo
  {pages} {042340} (\bibinfo {year} {2006})}\BibitemShut {NoStop}%
\bibitem [{\citenamefont {Cai}\ and\ \citenamefont
  {Scarani}(2009{\natexlab{a}})}]{Cai09}%
  \BibitemOpen
  \bibfield  {author} {\bibinfo {author} {\bibfnamefont {R.}~\bibnamefont
  {Cai}}\ and\ \bibinfo {author} {\bibfnamefont {V.}~\bibnamefont {Scarani}},\
  }\href {\doibase 10.1088/1367-2630/11/4/045024} {\bibfield  {journal}
  {\bibinfo  {journal} {New Journal of Physics}\ }\textbf {\bibinfo {volume}
  {11}},\ \bibinfo {pages} {045024} (\bibinfo {year}
  {2009}{\natexlab{a}})}\BibitemShut {NoStop}%
\bibitem [{\citenamefont {Sheridan}\ and\ \citenamefont
  {Scarani}(2010)}]{She10}%
  \BibitemOpen
  \bibfield  {author} {\bibinfo {author} {\bibfnamefont {L.}~\bibnamefont
  {Sheridan}}\ and\ \bibinfo {author} {\bibfnamefont {V.}~\bibnamefont
  {Scarani}},\ }\href {\doibase 10.1103/PhysRevA.82.030301} {\bibfield
  {journal} {\bibinfo  {journal} {Phys. Rev. A}\ }\textbf {\bibinfo {volume}
  {82}},\ \bibinfo {pages} {030301} (\bibinfo {year} {2010})}\BibitemShut
  {NoStop}%
\bibitem [{\citenamefont {Hasegawa}\ \emph {et~al.}()\citenamefont {Hasegawa},
  \citenamefont {Hayashi}, \citenamefont {Hiroshima}, \citenamefont {Tanaka},\
  and\ \citenamefont {Tomita}}]{Has07}%
  \BibitemOpen
  \bibfield  {author} {\bibinfo {author} {\bibfnamefont {J.}~\bibnamefont
  {Hasegawa}}, \bibinfo {author} {\bibfnamefont {M.}~\bibnamefont {Hayashi}},
  \bibinfo {author} {\bibfnamefont {T.}~\bibnamefont {Hiroshima}}, \bibinfo
  {author} {\bibfnamefont {A.}~\bibnamefont {Tanaka}}, \ and\ \bibinfo {author}
  {\bibfnamefont {A.}~\bibnamefont {Tomita}},\ }\href@noop {} {\enquote
  {\bibinfo {title} {Experimental decoy state quantum key distribution with
  unconditional security incorporating finite statistics},}\ }\Eprint
  {http://arxiv.org/abs/0705.3081} {arXiv:0705.3081 [quant-ph]} \BibitemShut
  {NoStop}%
\bibitem [{\citenamefont {K\"{o}nig}\ \emph {et~al.}(2009)\citenamefont
  {K\"{o}nig}, \citenamefont {Renner},\ and\ \citenamefont
  {Schaffner}}]{Sch08}%
  \BibitemOpen
  \bibfield  {author} {\bibinfo {author} {\bibfnamefont {R.}~\bibnamefont
  {K\"{o}nig}}, \bibinfo {author} {\bibfnamefont {R.}~\bibnamefont {Renner}}, \
  and\ \bibinfo {author} {\bibfnamefont {C.}~\bibnamefont {Schaffner}},\ }\href
  {\doibase 10.1109/TIT.2009.2025545} {\bibfield  {journal} {\bibinfo
  {journal} {IEEE Trans. Inf. Th.}\ }\textbf {\bibinfo {volume} {55}},\
  \bibinfo {pages} {4337} (\bibinfo {year} {2009})},\ \Eprint
  {http://arxiv.org/abs/0807.1338} {arXiv:0807.1338 [quant-ph]} \BibitemShut
  {NoStop}%
\bibitem [{\citenamefont {Bru\ss{}}(1998)}]{Bru98}%
  \BibitemOpen
  \bibfield  {author} {\bibinfo {author} {\bibfnamefont {D.}~\bibnamefont
  {Bru\ss{}}},\ }\href {\doibase 10.1103/PhysRevLett.81.3018} {\bibfield
  {journal} {\bibinfo  {journal} {Phys. Rev. Lett.}\ }\textbf {\bibinfo
  {volume} {81}},\ \bibinfo {pages} {3018} (\bibinfo {year}
  {1998})}\BibitemShut {NoStop}%
\bibitem [{\citenamefont {Bechmann-Pasquinucci}\ and\ \citenamefont
  {Gisin}(1999)}]{Bec99}%
  \BibitemOpen
  \bibfield  {author} {\bibinfo {author} {\bibfnamefont {H.}~\bibnamefont
  {Bechmann-Pasquinucci}}\ and\ \bibinfo {author} {\bibfnamefont
  {N.}~\bibnamefont {Gisin}},\ }\href {\doibase 10.1103/PhysRevA.59.4238}
  {\bibfield  {journal} {\bibinfo  {journal} {Phys. Rev. A}\ }\textbf {\bibinfo
  {volume} {59}},\ \bibinfo {pages} {4238} (\bibinfo {year}
  {1999})}\BibitemShut {NoStop}%
\bibitem [{\citenamefont {Bennett}\ \emph {et~al.}(1993)\citenamefont
  {Bennett}, \citenamefont {Brassard}, \citenamefont {Cr\'{e}peau},
  \citenamefont {Jozsa}, \citenamefont {Peres},\ and\ \citenamefont
  {Wootters}}]{Ben93}%
  \BibitemOpen
  \bibfield  {author} {\bibinfo {author} {\bibfnamefont {C.~H.}\ \bibnamefont
  {Bennett}}, \bibinfo {author} {\bibfnamefont {G.}~\bibnamefont {Brassard}},
  \bibinfo {author} {\bibfnamefont {C.}~\bibnamefont {Cr\'{e}peau}}, \bibinfo
  {author} {\bibfnamefont {R.}~\bibnamefont {Jozsa}}, \bibinfo {author}
  {\bibfnamefont {A.}~\bibnamefont {Peres}}, \ and\ \bibinfo {author}
  {\bibfnamefont {W.~K.}\ \bibnamefont {Wootters}},\ }\href {\doibase
  10.1103/PhysRevLett.70.1895} {\bibfield  {journal} {\bibinfo  {journal}
  {Phys. Rev. Lett.}\ }\textbf {\bibinfo {volume} {70}},\ \bibinfo {pages}
  {1895} (\bibinfo {year} {1993})}\BibitemShut {NoStop}%
\bibitem [{\citenamefont {Scarani}\ and\ \citenamefont
  {Renner}(2008{\natexlab{b}})}]{Sca08}%
  \BibitemOpen
  \bibfield  {author} {\bibinfo {author} {\bibfnamefont {V.}~\bibnamefont
  {Scarani}}\ and\ \bibinfo {author} {\bibfnamefont {R.}~\bibnamefont
  {Renner}},\ }in\ \href {\doibase 10.1007/978-3-540-89304-2_8} {\emph
  {\bibinfo {booktitle} {Theory of Quantum Computation, Communication, and
  Cryptography}}},\ \bibinfo {series} {Lecture Notes in Computer Science},
  Vol.\ \bibinfo {volume} {5106},\ \bibinfo {editor} {edited by\ \bibinfo
  {editor} {\bibfnamefont {Y.}~\bibnamefont {Kawano}}\ and\ \bibinfo {editor}
  {\bibfnamefont {M.}~\bibnamefont {Mosca}}}\ (\bibinfo  {publisher} {Springer
  Berlin / Heidelberg},\ \bibinfo {year} {2008})\ pp.\ \bibinfo {pages}
  {83--95}\BibitemShut {NoStop}%
\bibitem [{Note1()}]{Note1}%
  \BibitemOpen
  \bibinfo {note} {The formula in \cite {Sca08,Sca08a,Cai09} was corrected in
  an erratum \cite {Cai09err}. The formula in Eq.~\protect \textup {\hbox
  {\mathsurround \z@ \protect \normalfont (\ignorespaces \ref {eq:zeta}\unskip
  \@@italiccorr )}} can be obtained by multiplying the corresponding formula in
  \cite {Cai09err} by $\protect \frac {1}{2}$.}\BibitemShut {Stop}%
\bibitem [{\citenamefont {Biham}\ and\ \citenamefont {Mor}(1997)}]{Bih97}%
  \BibitemOpen
  \bibfield  {author} {\bibinfo {author} {\bibfnamefont {E.}~\bibnamefont
  {Biham}}\ and\ \bibinfo {author} {\bibfnamefont {T.}~\bibnamefont {Mor}},\
  }\href {\doibase 10.1103/PhysRevLett.78.2256} {\bibfield  {journal} {\bibinfo
   {journal} {Phys.\ Rev.\ Lett.}\ }\textbf {\bibinfo {volume} {78}},\ \bibinfo
  {pages} {2256} (\bibinfo {year} {1997})}\BibitemShut {NoStop}%
\bibitem [{\citenamefont {Holevo}(1978)}]{Hol78}%
  \BibitemOpen
  \bibfield  {author} {\bibinfo {author} {\bibfnamefont {A.}~\bibnamefont
  {Holevo}},\ }\href@noop {} {\bibfield  {journal} {\bibinfo  {journal}
  {Theor.\ Probab.\ Appl.}\ }\textbf {\bibinfo {volume} {23}},\ \bibinfo
  {pages} {411} (\bibinfo {year} {1978})}\BibitemShut {NoStop}%
\bibitem [{\citenamefont {Hausladen}\ and\ \citenamefont
  {Wootters}(1994)}]{Hau94}%
  \BibitemOpen
  \bibfield  {author} {\bibinfo {author} {\bibfnamefont {P.}~\bibnamefont
  {Hausladen}}\ and\ \bibinfo {author} {\bibfnamefont {W.}~\bibnamefont
  {Wootters}},\ }\href {\doibase 10.1080/09500349414552221} {\bibfield
  {journal} {\bibinfo  {journal} {J.\ Mod.\ Opt.}\ }\textbf {\bibinfo {volume}
  {41}},\ \bibinfo {pages} {2385} (\bibinfo {year} {1994})}\BibitemShut
  {NoStop}%
\bibitem [{\citenamefont {Ban}\ \emph {et~al.}(1997)\citenamefont {Ban},
  \citenamefont {Kurukowa}, \citenamefont {Momose},\ and\ \citenamefont
  {Hirota}}]{Ban97}%
  \BibitemOpen
  \bibfield  {author} {\bibinfo {author} {\bibfnamefont {M.}~\bibnamefont
  {Ban}}, \bibinfo {author} {\bibfnamefont {K.}~\bibnamefont {Kurukowa}},
  \bibinfo {author} {\bibfnamefont {R.}~\bibnamefont {Momose}}, \ and\ \bibinfo
  {author} {\bibfnamefont {O.}~\bibnamefont {Hirota}},\ }\href {\doibase
  10.1007/BF02435921} {\bibfield  {journal} {\bibinfo  {journal} {Int.\ J.\
  Theor.\ Phys.}\ }\textbf {\bibinfo {volume} {36}},\ \bibinfo {pages} {1269}
  (\bibinfo {year} {1997})}\BibitemShut {NoStop}%
\bibitem [{\citenamefont {Eldar}\ and\ \citenamefont {Forney}(2001)}]{Eld01}%
  \BibitemOpen
  \bibfield  {author} {\bibinfo {author} {\bibfnamefont {Y.}~\bibnamefont
  {Eldar}}\ and\ \bibinfo {author} {\bibfnamefont {G.}~\bibnamefont {Forney}},\
  }\href {\doibase 10.1109/18.915636} {\bibfield  {journal} {\bibinfo
  {journal} {IEEE Trans.\ Inf.\ Theory}\ }\textbf {\bibinfo {volume} {47}},\
  \bibinfo {pages} {858} (\bibinfo {year} {2001})}\BibitemShut {NoStop}%
\bibitem [{\citenamefont {Chefles}(2000)}]{Che00}%
  \BibitemOpen
  \bibfield  {author} {\bibinfo {author} {\bibfnamefont {A.}~\bibnamefont
  {Chefles}},\ }\href {\doibase 10.1080/00107510010002599} {\bibfield
  {journal} {\bibinfo  {journal} {Contemp.\ Phys.}\ }\textbf {\bibinfo {volume}
  {41}},\ \bibinfo {pages} {401} (\bibinfo {year} {2000})}\BibitemShut
  {NoStop}%
\bibitem [{\citenamefont {Barnett}(2001)}]{Bar01}%
  \BibitemOpen
  \bibfield  {author} {\bibinfo {author} {\bibfnamefont {S.}~\bibnamefont
  {Barnett}},\ }\href {\doibase 10.1103/PhysRevA.64.030303} {\bibfield
  {journal} {\bibinfo  {journal} {Phys.\ Rev.\ A}\ }\textbf {\bibinfo {volume}
  {64}},\ \bibinfo {pages} {030303} (\bibinfo {year} {2001})}\BibitemShut
  {NoStop}%
\bibitem [{\citenamefont {Holevo}(1973)}]{Hol73}%
  \BibitemOpen
  \bibfield  {author} {\bibinfo {author} {\bibfnamefont {A.~S.}\ \bibnamefont
  {Holevo}},\ }\href@noop {} {\bibfield  {journal} {\bibinfo  {journal}
  {Journal of Multivariate Analysis}\ }\textbf {\bibinfo {volume} {3}},\
  \bibinfo {pages} {337} (\bibinfo {year} {1973})}\BibitemShut {NoStop}%
\bibitem [{\citenamefont {Yuen}\ \emph {et~al.}(1975)\citenamefont {Yuen},
  \citenamefont {Kennedy},\ and\ \citenamefont {Lax}}]{Yuen75}%
  \BibitemOpen
  \bibfield  {author} {\bibinfo {author} {\bibfnamefont {H.}~\bibnamefont
  {Yuen}}, \bibinfo {author} {\bibfnamefont {R.}~\bibnamefont {Kennedy}}, \
  and\ \bibinfo {author} {\bibfnamefont {M.}~\bibnamefont {Lax}},\ }\href@noop
  {} {\bibfield  {journal} {\bibinfo  {journal} {IEEE Trans. on Inform.
  Theor.}\ }\textbf {\bibinfo {volume} {21}},\ \bibinfo {pages} {125} (\bibinfo
  {year} {1975})}\BibitemShut {NoStop}%
\bibitem [{\citenamefont {Helstr\o{}m}(1976)}]{Hel76}%
  \BibitemOpen
  \bibfield  {author} {\bibinfo {author} {\bibfnamefont {C.}~\bibnamefont
  {Helstr\o{}m}},\ }\href@noop {} {\emph {\bibinfo {title} {Quantum Detection
  and Estimation Theory}}}\ (\bibinfo  {publisher} {Academic, New York},\
  \bibinfo {year} {1976})\BibitemShut {NoStop}%
\bibitem [{\citenamefont {Herzog}\ and\ \citenamefont {Bergou}(2004)}]{Her04}%
  \BibitemOpen
  \bibfield  {author} {\bibinfo {author} {\bibfnamefont {U.}~\bibnamefont
  {Herzog}}\ and\ \bibinfo {author} {\bibfnamefont {J.}~\bibnamefont
  {Bergou}},\ }\href {\doibase 10.1103/PhysRevA.70.022302} {\bibfield
  {journal} {\bibinfo  {journal} {Phys. Rev. A}\ }\textbf {\bibinfo {volume}
  {70}},\ \bibinfo {pages} {022302} (\bibinfo {year} {2004})}\BibitemShut
  {NoStop}%
\bibitem [{\citenamefont {Qiu}\ and\ \citenamefont {Li}(2010)}]{Qiu10}%
  \BibitemOpen
  \bibfield  {author} {\bibinfo {author} {\bibfnamefont {D.}~\bibnamefont
  {Qiu}}\ and\ \bibinfo {author} {\bibfnamefont {L.}~\bibnamefont {Li}},\
  }\href {\doibase 10.1103/PhysRevA.81.042329} {\bibfield  {journal} {\bibinfo
  {journal} {Phys.\ Rev.\ A}\ }\textbf {\bibinfo {volume} {81}},\ \bibinfo
  {pages} {042329} (\bibinfo {year} {2010})}\BibitemShut {NoStop}%
\bibitem [{\citenamefont {Bechmann-Pasquinucci}\ and\ \citenamefont
  {Peres}(2000)}]{Bechmann2000}%
  \BibitemOpen
  \bibfield  {author} {\bibinfo {author} {\bibfnamefont {H.}~\bibnamefont
  {Bechmann-Pasquinucci}}\ and\ \bibinfo {author} {\bibfnamefont
  {A.}~\bibnamefont {Peres}},\ }\href {\doibase 10.1103/PhysRevLett.85.3313}
  {\bibfield  {journal} {\bibinfo  {journal} {Phys. Rev. Lett.}\ }\textbf
  {\bibinfo {volume} {85}},\ \bibinfo {pages} {3313} (\bibinfo {year}
  {2000})}\BibitemShut {NoStop}%
\bibitem [{\citenamefont {Bru\ss{}}\ and\ \citenamefont
  {Macchiavello}(2002)}]{Bru02}%
  \BibitemOpen
  \bibfield  {author} {\bibinfo {author} {\bibfnamefont {D.}~\bibnamefont
  {Bru\ss{}}}\ and\ \bibinfo {author} {\bibfnamefont {C.}~\bibnamefont
  {Macchiavello}},\ }\href {\doibase 10.1103/PhysRevLett.88.127901} {\bibfield
  {journal} {\bibinfo  {journal} {Phys. Rev. Lett.}\ }\textbf {\bibinfo
  {volume} {88}},\ \bibinfo {pages} {127901} (\bibinfo {year}
  {2002})}\BibitemShut {NoStop}%
\bibitem [{\citenamefont {Cerf}\ \emph {et~al.}(2002)\citenamefont {Cerf},
  \citenamefont {Bourennane}, \citenamefont {Karlsson},\ and\ \citenamefont
  {Gisin}}]{Cer02}%
  \BibitemOpen
  \bibfield  {author} {\bibinfo {author} {\bibfnamefont {N.}~\bibnamefont
  {Cerf}}, \bibinfo {author} {\bibfnamefont {M.}~\bibnamefont {Bourennane}},
  \bibinfo {author} {\bibfnamefont {A.}~\bibnamefont {Karlsson}}, \ and\
  \bibinfo {author} {\bibfnamefont {N.}~\bibnamefont {Gisin}},\ }\href
  {\doibase 10.1103/PhysRevLett.88.127902} {\bibfield  {journal} {\bibinfo
  {journal} {Phys. Rev. Lett.}\ }\textbf {\bibinfo {volume} {88}},\ \bibinfo
  {pages} {127902} (\bibinfo {year} {2002})}\BibitemShut {NoStop}%
\bibitem [{\citenamefont {Bru\ss{}}\ \emph {et~al.}(2003)\citenamefont
  {Bru\ss{}} \emph {et~al.}}]{Bru03}%
  \BibitemOpen
  \bibfield  {author} {\bibinfo {author} {\bibfnamefont {D.}~\bibnamefont
  {Bru\ss{}}} \emph {et~al.},\ }\href {\doibase 10.1103/PhysRevLett.91.097901}
  {\bibfield  {journal} {\bibinfo  {journal} {Phys.\ Rev.\ Lett.}\ }\textbf
  {\bibinfo {volume} {91}},\ \bibinfo {pages} {097901} (\bibinfo {year}
  {2003})}\BibitemShut {NoStop}%
\bibitem [{\citenamefont {Liang}\ \emph {et~al.}(2003)\citenamefont {Liang}
  \emph {et~al.}}]{Lia03}%
  \BibitemOpen
  \bibfield  {author} {\bibinfo {author} {\bibfnamefont {Y.}~\bibnamefont
  {Liang}} \emph {et~al.},\ }\href {\doibase 10.1103/PhysRevA.68.022324}
  {\bibfield  {journal} {\bibinfo  {journal} {Phys.\ Rev.\ A}\ }\textbf
  {\bibinfo {volume} {68}},\ \bibinfo {pages} {022324} (\bibinfo {year}
  {2003})}\BibitemShut {NoStop}%
\bibitem [{\citenamefont {Kaszlikowski}\ \emph {et~al.}(2004)\citenamefont
  {Kaszlikowski} \emph {et~al.}}]{Kas04}%
  \BibitemOpen
  \bibfield  {author} {\bibinfo {author} {\bibfnamefont {D.}~\bibnamefont
  {Kaszlikowski}} \emph {et~al.},\ }\href {\doibase 10.1103/PhysRevA.70.032306}
  {\bibfield  {journal} {\bibinfo  {journal} {Phys.\ Rev.\ A}\ }\textbf
  {\bibinfo {volume} {70}},\ \bibinfo {pages} {032306} (\bibinfo {year}
  {2004})}\BibitemShut {NoStop}%
\bibitem [{\citenamefont {Scarani}(2010)}]{Scarani2010}%
  \BibitemOpen
  \bibfield  {author} {\bibinfo {author} {\bibfnamefont {V.}~\bibnamefont
  {Scarani}},\ }in\ \href@noop {} {\emph {\bibinfo {booktitle} {Quantum
  Cryptography and Computing}}},\ \bibinfo {series} {NATO Science for Peace and
  Security Series - D: Information and Communication Security}, Vol.~\bibinfo
  {volume} {26},\ \bibinfo {editor} {edited by\ \bibinfo {editor}
  {\bibfnamefont {R.}~\bibnamefont {Horodecki}}, \bibinfo {editor}
  {\bibfnamefont {S.~Y.}\ \bibnamefont {Kilin}}, \ and\ \bibinfo {editor}
  {\bibfnamefont {J.}~\bibnamefont {Kowalik}}}\ (\bibinfo {year} {2010})\ pp.\
  \bibinfo {pages} {76--82},\ \Eprint {http://arxiv.org/abs/arXiv:1010.0521
  [quant-ph]} {arXiv:1010.0521 [quant-ph]} \BibitemShut {NoStop}%
\bibitem [{\citenamefont {Sheridan}\ and\ \citenamefont
  {Scarani}(2011)}]{Sheridan2011}%
  \BibitemOpen
  \bibfield  {author} {\bibinfo {author} {\bibfnamefont {L.}~\bibnamefont
  {Sheridan}}\ and\ \bibinfo {author} {\bibfnamefont {V.}~\bibnamefont
  {Scarani}},\ }\href {\doibase 10.1103/PhysRevA.83.039901} {\bibfield
  {journal} {\bibinfo  {journal} {Phys. Rev. A}\ }\textbf {\bibinfo {volume}
  {83}},\ \bibinfo {pages} {039901} (\bibinfo {year} {2011})}\BibitemShut
  {NoStop}%
\bibitem [{\citenamefont {Durt}\ \emph {et~al.}()\citenamefont {Durt},
  \citenamefont {Englert}, \citenamefont {Bengtsson},\ and\ \citenamefont
  {\.{Z}yczkowski}}]{Dur10}%
  \BibitemOpen
  \bibfield  {author} {\bibinfo {author} {\bibfnamefont {T.}~\bibnamefont
  {Durt}}, \bibinfo {author} {\bibfnamefont {B.-G.}\ \bibnamefont {Englert}},
  \bibinfo {author} {\bibfnamefont {I.}~\bibnamefont {Bengtsson}}, \ and\
  \bibinfo {author} {\bibfnamefont {K.}~\bibnamefont {\.{Z}yczkowski}},\
  }\href@noop {} {\enquote {\bibinfo {title} {On mutually unbiased bases},}\
  }\Eprint {http://arxiv.org/abs/1004.3348} {arXiv:1004.3348 [quant-ph]}
  \BibitemShut {NoStop}%
\bibitem [{\citenamefont {Cover}\ and\ \citenamefont {Thomas}(1991)}]{Cov91}%
  \BibitemOpen
  \bibfield  {author} {\bibinfo {author} {\bibfnamefont {T.}~\bibnamefont
  {Cover}}\ and\ \bibinfo {author} {\bibfnamefont {J.}~\bibnamefont {Thomas}},\
  }\href@noop {} {\emph {\bibinfo {title} {Elements of Information Theory}}}\
  (\bibinfo  {publisher} {Wiley, New York},\ \bibinfo {year}
  {1991})\BibitemShut {NoStop}%
\bibitem [{\citenamefont {Cai}\ and\ \citenamefont
  {Scarani}(2009{\natexlab{b}})}]{Cai09err}%
  \BibitemOpen
  \bibfield  {author} {\bibinfo {author} {\bibfnamefont {R.~Y.~Q.}\
  \bibnamefont {Cai}}\ and\ \bibinfo {author} {\bibfnamefont {V.}~\bibnamefont
  {Scarani}},\ }\href {\doibase 10.1088/1367-2630/11/10/109801} {\bibfield
  {journal} {\bibinfo  {journal} {New Journal of Physics}\ }\textbf {\bibinfo
  {volume} {11}},\ \bibinfo {pages} {109801} (\bibinfo {year}
  {2009}{\natexlab{b}})}\BibitemShut {NoStop}%
\end{thebibliography}%

\end{document}